\documentclass[pra, 10pt, twocolumn, tightenlines, longbibliography, aps,superscriptaddress,nofootinbib]{revtex4-2}

\usepackage{comment}

\usepackage[dvips]{graphics}
\usepackage{color}
\usepackage{float}
\usepackage{natbib}
\usepackage[dvipsnames]{xcolor}
\setcitestyle{numbers}
\setcitestyle{square}
\usepackage{graphicx}
\usepackage{bm}
\usepackage{amssymb}
\usepackage{amsmath}
\usepackage{mathtools}
\usepackage{dcolumn}  
\usepackage[colorlinks, linkcolor=Blue,citecolor=Blue,urlcolor=Blue]{hyperref}

\usepackage[all]{hypcap} 
\usepackage{physics}
\usepackage{enumerate}
\usepackage{braket}       
\usepackage{overpic}
\usepackage{amsfonts}
\usepackage{dsfont}
\usepackage[mathscr]{eucal}
\usepackage{hyperref}
\usepackage{setspace}
\usepackage{lipsum} 

\newcommand{\magenta}[1]{\textcolor{magenta}{#1}}

\usepackage[normalem]{ulem} %% REMOVE ME!

\begin{document}
% \title{Josephson Oscillation detection using Cavity Optomechanics }
%\title{Nondestructive Detection of Josephson Oscillations using Cavity Optomechanics}
\title{Nonreciprocal Josephson Dynamics of a $\mathcal{PT}$-Symmetric Atomtronic SQUID}
\author{Nalinikanta Pradhan}
\affiliation{Department of Physics, Indian Institute of Technology Guwahati, Guwahati, Assam, 781039, India}

\author{Rina Kanamoto}
\affiliation{Department of Physics, Meiji University, Kawasaki, Kanagawa 214-8571, Japan}

\author{M. Bhattacharya}
\affiliation{School of Physics and Astronomy, Rochester Institute of Technology, 84 Lomb Memorial Drive, Rochester, New York 14623, USA}
\author{Pankaj Kumar Mishra}
\affiliation{Department of Physics, Indian Institute of Technology Guwahati, Guwahati, Assam, 781039, India}

\date{\today}

\begin{abstract}

We investigate the non-Hermitian Josephson dynamics in an atomtronic quantum interference device (AQUID, or atomtronic SQUID) formed by a ring Bose–Einstein condensate with a balanced atomic loss-gain region and interrupted by two moving optical barriers. The spatially balanced gain–loss profile establishes a $\mathcal{PT}$-symmetric potential while breaking spatial inversion symmetry, providing the essential ingredients for nonreciprocal Josephson transport. The resulting non-Hermitian dynamics are monitored nondestructively via the phase-quadrature spectrum of an optomechanical cavity driven by Laguerre-Gaussian beams that couple to the condensate. We demonstrate that with increasing gain–loss strength ($\Gamma_{\mathrm{local}}$), the $\mathcal{PT}$-symmetry-induced collective mode softens and approaches an exceptional point (verified using the Bogoliubov-de Gennes (BdG) analysis), while its coherent population-exchange dynamics give rise to an additional characteristic splitting in the cavity-output spectrum. For linear translational motion of the optical barriers, the introduction of non-Hermiticity provides an additional control parameter for tuning the critical current of the circuit. When the barriers undergo simultaneous translation and rotation, the rotation-induced phase modifies the interference between counter-propagating currents, providing an additional means of controlling the nonreciprocal response. Under periodic barrier modulation, we further demonstrate that the Shapiro steps in the current-voltage curve are more pronounced due to the intrinsic $\mathcal{PT}$-symmetric collective mode. Finally, we propose an alternative Hermitian realization based on a tilted optical-lattice potential, offering an experimentally accessible route to engineering biased transport and population-transfer dynamics. Our results establish $\mathcal{PT}$-symmetric atomtronic SQUIDs as a versatile platform for controlling nonreciprocal Josephson transport and collective excitations in driven quantum fluids.

\end{abstract}

%\flushbottom

\maketitle
%%%%%%%%%%%%%%%%%%%%%%%%%%%%%%%%%%%%%%%%%%%%%%%%%%%%%%%%%%%%%%%

\section{Introduction}
Atomtronic circuits provide a versatile platform for realizing quantum analogues of electronic devices, with superfluid Bose–Einstein condensates (BECs) serving as coherent carriers of matter-wave currents~\cite{Amico2021roadmap, Amico_2017Focus}. Among these systems, atomtronic superconducting quantum interference devices (SQUIDs), consisting of a ring-shaped BEC interrupted by two optical barriers, provide a highly controllable platform for studying Josephson tunneling, quantum interference, and coherent matter-wave transport~\cite{RyuPRL2013, PhysRevA.97.013633, ryu2020quantum}. The two optical barriers act as Josephson junctions, and the resulting dynamics are governed by the interplay of tunneling, interatomic interactions, phase winding, and the external trapping potential. In particular, the motion of the barriers provides a direct means of controlling transport across the junctions. When the barrier velocity remains below a critical threshold value, the system exhibits a dc Josephson response, whereas, above the critical velocity the dynamics enter an ac Josephson regime characterized by phase slips and sustained population transfer~\cite{Giovanazzi2000, RyuPRL2013}. The critical barrier velocity thus plays the role of the critical current in a conventional Josephson junction~\cite{Giovanazzi2000}. The ring geometry further enables quantum interference between the currents flowing through the two junctions, giving rise to a rich variety of phenomena, including phase slips of persistent currents~\cite{WrightPRL2013,pezze2024stabilizing,Kiehn2022,chatterjee2026spinorcondensatepersistentcurrents}, current oscillations~\cite{Polo2019Oscillations}, dc-ac Josephson effect \cite{RyuPRL2013,Cataldo2014Dynamics,Jendrzejewski2014Resistive,Cataldo2020}, superfluid qubits~\cite{solenov2010cold,amico2014superfluidqubit}, interference of circulating currents~\cite{aghamalyan2015coherent, readout2018,ryu2020quantum,Kiehn2022,pradhan2025fractional}, integer and fractional Shapiro steps~\cite{bernhart2024observation, pradhan2025fractional}, Josephson vortices~\cite{borysenko2024acceleration,Singh2026Controlled}, Fraunhofer-like interference patterns~\cite{geier2026fraunhofer}, and the Josephson diode effect~\cite{Pradhan2026diode}.

An additional route for controlling Josephson dynamics in atomtronic circuits is provided by non-Hermitian engineering, in which spatially balanced gain and loss can be introduced to realize an effective $\mathcal{PT}$-symmetric system~\cite{Cartarius2012Model,Fabian2014Coupling,ppvl-5lmg,Fortanier2014Dipolar,Lhle2014STABILITY,Dennis2016Bose,Dmitry2016Stationary,Lukas2017Vortices,Daniel2018Haag}. Unlike conventional Hermitian systems, $\mathcal{PT}$-symmetric systems can undergo a transition between regimes of real and complex eigenvalues, accompanied by the formation of exceptional points, where eigenvalues and their corresponding eigenmodes coalesce~\cite{ElGanainy2018PTReview,Ozdemir2019ExceptionalPoints, Miri2019ExceptionalPoints}. Incorporating such non-Hermitian coupling into an atomtronic SQUID could provide a natural framework for exploring the interplay between biased Josephson transport, quantum interference, and collective non-Hermitian dynamics~\cite{Takasu2020}. In particular, a controlled energy bias between different regions of the condensate can be generated using a tilted-well potential, thereby modifying the population and phase evolution of the system~\cite{Kreibich2013Hermitian,Kreibich2014Realizing,Kreibich2016Tilted,Kolovsky2010}. A spatially engineered tilted-well configuration in the presence of optical barriers can therefore provide an experimentally accessible route to controlling biased population transfer while retaining an overall Hermitian description. Such a realization is particularly relevant for atomtronic circuits, where controlled transport and population redistribution are essential ingredients for designing functional quantum devices~\cite{Amico2021roadmap, AmicoRMP2022}.

% Despite significant progress in Hermitian atomtronic SQUIDs, how balanced gain and loss reshape their collective modes, critical transport, and quantum interference remains largely unexplored. In contrast to conventional Josephson dynamics, governed by tunneling, interactions, phase bias, and barrier motion, $\mathcal{PT}$-symmetric coupling provides an additional channel for coherent population exchange and phase evolution. This raises a central question: can non-Hermiticity actively engineer collective Josephson dynamics and non-reciprocal matter-wave transport?

In this paper, we establish a connection between $\mathcal{PT}$-symmetric collective dynamics and non-reciprocal Josephson transport in an atomtronic SQUID. Balanced gain and loss excite an additional collective mode alongside the conventional Josephson plasma mode, whose frequency softens toward an exceptional point as the gain–loss strength increases. The associated coherent population exchange produces an additional splitting in the cavity-output spectrum, providing a direct spectroscopic signature of the $\mathcal{PT}$ dynamics~\cite{ElGanainy2018PTReview}. We further show that the non-Hermitian coupling leads to direction-dependent critical barrier velocities (critical currents), realizing an atomtronic diode and asymmetric Shapiro-like steps. Simultaneous translation and rotation of the barriers provide additional control over quantum interference, critical velocity, synchronization, and non-reciprocal response, while periodic modulation enables resonant and off-resonant synchronization with the intrinsic $\mathcal{PT}$ mode. These dynamics are characterized using coupled stochastic Gross–Pitaevskii and quantum Langevin equations, complemented by Bogoliubov–de Gennes analysis, and can be probed nondestructively through the power spectrum of the phase-quadrature of the cavity output field~\cite{pradhan2024cavity,Pradhan2026diode}. Finally, we demonstrate an alternative Hermitian realization using a tilted optical-lattice potential, which reproduces the essential biased population-transfer dynamics without requiring an explicitly complex potential~\cite{Lai2016SourceSink}. Overall, our work provides a unified framework for engineering and probing non-Hermitian Josephson transport, collective dynamics, and quantum interference in atomtronic circuits.

Our manuscript is organized as follows. In Sec.\ref{sec:model}, we introduce the theoretical model, followed by the $\mathcal{PT}$-symmetric potential and optical-barrier configuration in Sec.\ref{sec:PT}. Sections\ref{subsec:linear_dc_ac}, ~\ref{sec:Shapiro}, and \ref{subsec:linear_rotation} present the barrier-driven Josephson dynamics, including the dc–ac Josephson transition, Shapiro steps, quantum interference, and diode behaviour, while Sec.\ref{sec:titltedLattice} considers the corresponding dynamics in a tilted optical lattice. Finally, in Sec.\ref{sec:conclusion}, we summarize our results and outline possible directions for future work. The collective excitations are analyzed within the Bogoliubov–de Gennes framework in the Appendix~\ref{sec:BdG}. 

\section{Theoretical Mean-field Model}
\label{sec:model}
We consider a ring-trapped Bose-Einstein condensate subject to two optical barriers and a spatially structured $\mathcal{PT}$-symmetric potential. To realize a nondestructive probe of the resulting atomtronic dynamics, the condensate is dispersively coupled to a single cavity mode \cite{BrenneckeScience2008, KumarPRL2021}. In this case, the atomtronic circuit is placed at the centre of an optical cavity, which is driven by two Laguerre-Gaussian beams of opposite charge $\pm \ell$ generating an optical lattice potential. In the context of a persistent current with winding $L_p$, when the condensate matter waves in the ring experience this lattice, they Bragg diffract and generate new side modes $L_p \pm 2\ell$. These side modes act as mechanical oscillators in cavity optomechanics and modulate the cavity resonance frequency. Subsequently, homodyne detection of the cavity output field yields the side-mode frequencies and the corresponding winding number of the persistent current \cite{KumarPRL2021,pradhan2024ring,pradhan2024cavity, pradhan2025AndreevBashkin}. However, along with the persistent current, if any kind of particle oscillations (i.e., Josephson oscillations) are generated in the circuit, the corresponding oscillation frequencies are manifested as the splitting of the side-mode peaks in the cavity output spectrum \cite{pradhan2025fractional,Pradhan2026diode}. 

In the current study, the above-discussed experimental setup is simulated by numerically solving the coupled stochastic Gross-Pitaevskii equation and the classical Langevin equation for the light-field amplitude. With a mean field and one-dimensional approximation (frozen atomic motion in radial and transverse directions in the thin ring) \cite{KumarPRL2021}, these equations can be written in a dimensionless form as \cite{pradhan2024cavity,pradhan2024ring,stoof2001dynamics, Snizhko2016Stochastic, das2012winding}
\begin{equation}
\begin{aligned}
    &(i-\Gamma)\frac{d\psi}{d\tau} = \biggl[\left( -i \frac{d}{d\phi} - \Omega'\right)^2 + \frac{U_0}{\omega_\beta}  |\alpha(\tau)|^2 \cos^2\left({\ell\phi}\right) \\ & + V_{\mathrm{bar}} (\phi,\tau) + V_{\mathrm{PT}} (\phi)   + \mathcal{G} | \psi (\phi,\tau)|^2 \biggr] \psi + \xi(\phi,\tau),
    \label{Eq:bec}    
\end{aligned}
\end{equation}
\begin{equation}
\begin{aligned}
    i\frac{d\alpha}{d\tau} = \biggl\{ - \biggl[\Delta_c &- U_0 \langle \cos^{2}\left(\ell\phi\right)\rangle_{\tau} + i \frac{\gamma_{0}}{2}\biggr] \alpha  + i\eta_{\mathrm{p}} \biggr\} \omega_\beta^{-1} \\ &+ i \sqrt{\gamma_0} \omega_\beta^{-1} \alpha_{in}(\tau),
    \label{Eq:cavity}
\end{aligned}
\end{equation}
with scaling energy as $\hbar \omega_{\beta} = \hbar^2 / 2 m R_0 ^2$ and time as $\tau = \omega_\beta t\;$. Here $m$ is the mass of the $^{23}$Na atom and $R_0$ the ring trap radius, which also serves as the unit of length. Here, $\psi(\phi,\tau)$ denotes the condensate wave-function, $\phi$ is the scaled azimuthal coordinate, and $\Omega' = \Omega/2\omega_\beta$ is the dimensionless rotation frequency. The parameter $\mathcal{G} = g/ \hbar\omega_\beta$ characterises the scaled 1D contact-interaction strength, while $\Gamma$ models the phenomenological damping in the condensate and $\xi(\phi,\tau)$ denotes the associated thermal fluctuations of the condensate at temperature $T$ \cite{das2012winding,pradhan2024ring, pradhan2025fractional}. The intracavity field $\alpha(\tau)$ generates a dynamic optical lattice potential with spatial periodicity set by the mode index $\ell$. Conversely, the condensate modifies the cavity resonance through the dispersive overlap $\langle\cos^2(\ell\phi)\rangle_{\tau}$, thereby establishing a nonlinear feedback between the atomic and photonic degrees of freedom. Here, $\Delta_c$ is the cavity-pump detuning, $U_0$ the single-atom dispersive coupling strength, $\gamma_0$ is the cavity-field decay rate, and the pump rate $\eta_{\mathrm{p}}=\sqrt{P_{\mathrm{in}}\gamma_{0}/\hbar\omega_{0}}$, which depends on the input power $P_{\mathrm{in}}$ and drive-cavity detuning $\Delta_{o}=\omega_{\mathrm{L}}-\omega_{0}$. The input field $\alpha_{\mathrm{in}}(\tau)$ accounts for fluctuations entering through the cavity input channel, and the output field from the cavity is related to the input field via the input-output relation $\alpha_{out} = -\alpha_{in} + \sqrt{\gamma_0}\alpha$ \cite{AspelmeyerRMP2014}. The thermal and shot noise terms in the stochastic equations are modelled by using delta-correlated white noise with zero mean, unit variance and have the correlation
\begin{align} 
\langle\xi(\phi,\tau) \, \xi ^ *(\phi',\tau')\rangle  &= \frac{2\,\Gamma \,k_B \,T}{\hbar \omega_\beta} \, \delta(\phi - \phi') \, \delta (\tau - \tau'), \\ 
\langle \alpha_{in}(\tau) \, \alpha_{in}^ *(\tau') \rangle  &= \omega_\beta \, \delta (\tau - \tau'),
\end{align}

The two localized optical barriers that serve as the Josephson junctions in the atomtronic circuit are modelled as \cite{Kiehn2022,KSgan2025josephson2D,pezze2024stabilizing, Pradhan2026diode}
\begin{equation}
    V_{\mathrm{bar}}(\phi,\tau) =
\sum_{j=1}^{2} 
V_j \exp\!\left[-2\left(\tfrac{\phi-\phi_{b_j}(\tau)}{\sigma_j}\right)^{2}\right].
\label{Eq:Vbar}
\end{equation}
Here, $V_j$, $\phi_{b_j}(\tau)$, and $\sigma_j$ denote, respectively, the height, centre position, and width of the $j$th barrier. The two optical barriers are then moved towards each other across the ring to study the Josephson transport phenomena, such as the dc-ac Josephson effect (linear motion), quantum interference of currents (linear motion in the rotating frame), and the Shapiro steps (periodically modulated barrier positions along with the linear motion). In the simulations, these motions are modelled as the time-dependent optical barrier potential, with their positions
\begin{equation}
     \phi_{b_{1,2}} = \pm (2 \pi f_{\mathrm{bar}} ) t \pm \phi_\mathrm{m} \, \sin(2 \pi f_{\mathrm{m}} t),
 \label{Eq:bar_pos}
\end{equation}
where $f_{\mathrm{bar}}$ is the barrier velocity, $\phi_\mathrm{m}$ is the amplitude of position modulation and $f_{\mathrm{m}}$ is the frequency of position modulation. These provide a controlled bias for investigating Josephson transport in the presence of rotation, dissipation, cavity-mediated feedback, and non-Hermitian gain-loss processes.
%--------------------------------------------------------------------
\begin{figure*}[!htp]
\begin{center}
\includegraphics[width= 1.0\linewidth]{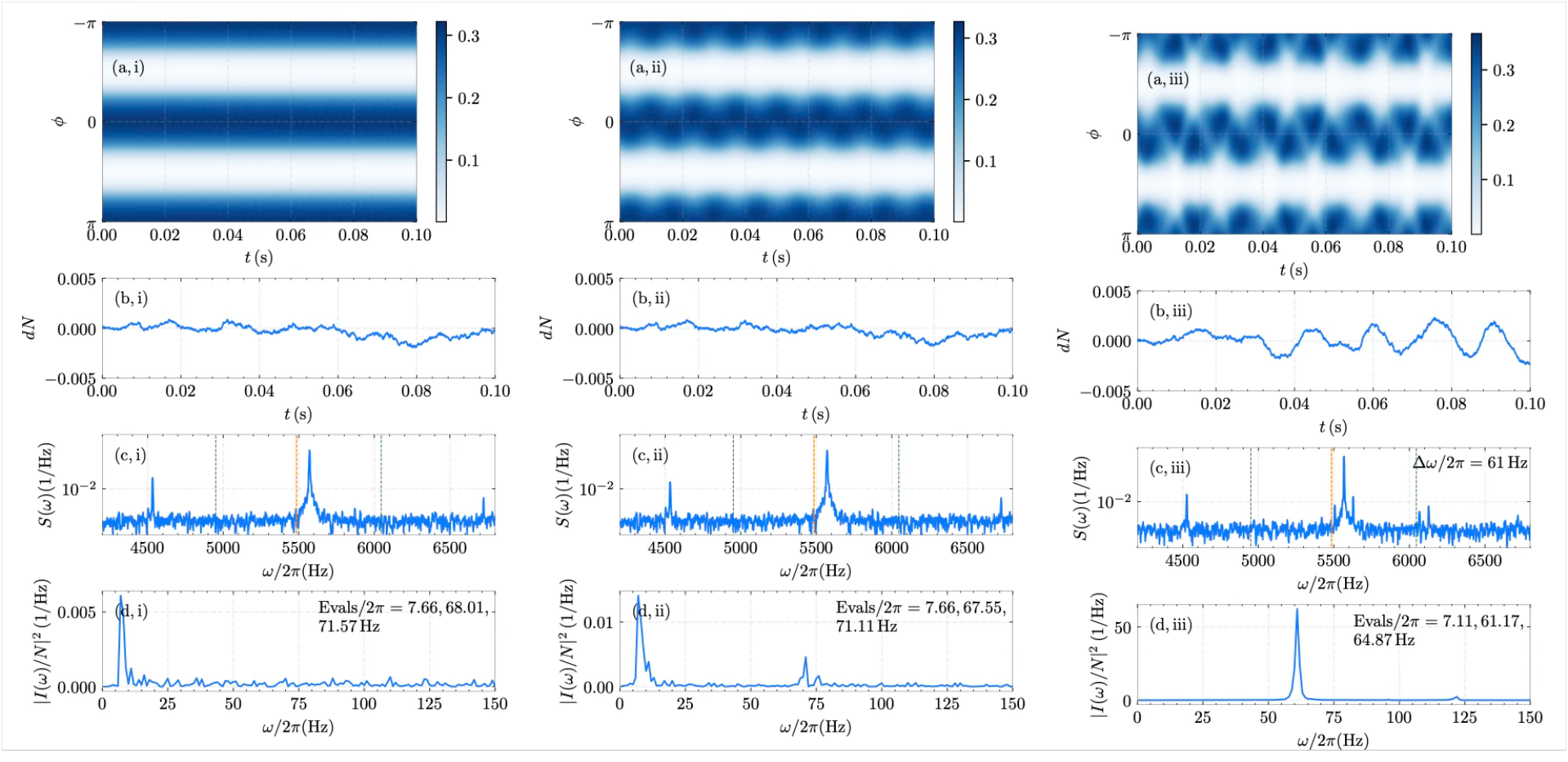} 

\end{center}
\vspace{-12pt}
\caption{Dynamics and cavity response of the atomtronic SQUID in the presence of the complex $\mathcal{PT}$-symmetric potential for $f_{\mathrm{bar}}=0$. Columns correspond to $\Gamma_{\mathrm{local}}=0.0$ (i), $0.5$ (ii), and $2.0$ (iii). (a) Temporal evolution of the condensate density, (b) population imbalance, (c) power spectrum of the phase quadrature of the cavity output field, and (d) power spectrum of the population imbalance. In (c), the vertical dashed lines indicate the side-mode frequencies of uniform-density persistent-current states with $L_p=0$ (orange) and $L_p=\pm1$ (green) for $\ell=10$. The eigenfrequencies in (d) are obtained from a Bogoliubov-de Gennes (BdG) analysis including the optical barriers and $\mathcal{PT}$-symmetric potential. See Sec. \ref{sec:BdG} for the details on the BdG analysis.  Parameters are $N=3700$, $R_0=4~\mu\mathrm{m}$, $V_0=17.5\mu_0=25~\mathrm{nK}$, $\sigma=0.7$, $\zeta=1.8~\mu\mathrm{m}$, $t_{\mathrm{bar}}=0.1~\mathrm{s}$, $t_{\mathrm{cav}}=1.0~\mathrm{s}$, $T=10~\mathrm{nK}$, $\omega_\rho=\omega_z=2\pi\times42\mathrm{Hz}$, $a_S = 2.5$ nm, $g = 2\hbar \omega_\rho a_s/R_0$, $P_{\mathrm{in}}=10~\mathrm{fW}$, $\Gamma=10^{-4}$, $U_0=2\pi\times212~\mathrm{Hz}$, $\widetilde{\Delta}=\Delta_0-U_0N/2=-2\pi\times173~\mathrm{Hz}$, $\Delta_a=2\pi\times4.7~\mathrm{GHz}$, $\omega_0=2\pi\times10^{15}\mathrm{Hz}$, $\gamma_0=2\pi\times2~\mathrm{MHz}$, and $\Omega'=0$.}
\label{fig:fig1}
\end{figure*}
 %-------------------------------------------------------
To introduce spatially balanced gain and loss, we consider an imaginary potential of the form \cite{ppvl-5lmg, Haag2018PTsymmetric} 
\begin{equation}
    V_{\mathrm{PT}} (\phi) = i \Gamma_{\mathrm{local}} \sin(\phi),
    \label{eq:PT_sin}
\end{equation}
where $\Gamma_{\mathrm{local}}$ characterises the strength of the local gain-loss of the atom from the condensate. The sinusoidal profile of Eq.~\eqref{eq:PT_sin} represents an idealized continuous angular modulation of the experimentally controllable loss and gain of particles using an external environment that exchanges particles with the condensate~\cite{Fabian2014Coupling}. Since the imaginary component of the complex potential (Eq.~\eqref{eq:PT_sin}) is antisymmetric under spatial inversion, and $
V_{\mathrm{PT}}(-\phi)
=
V_{\mathrm{PT}}^{*}(\phi)$,
the required $\mathcal{PT}$-symmetry condition in the atomtronic SQUID is established. Positive and negative portions of the modulation correspond, respectively, to effective gain and loss of atoms from the condensate in the non-Hermitian mean-field description. The resulting balanced gain-loss landscape therefore provides a controlled extension of the atomtronic SQUID and allows us to investigate the interplay between Josephson transport, rotation, cavity-mediated feedback, and $\mathcal{PT}$-symmetric non-Hermitian dynamics~\cite{Lunt2017}.

In the context of open quantum systems, such spatially structured imaginary potentials can be interpreted as effective one-body gain and loss terms arising from spatially controlled atomic dissipation. In ultracold-atom experiments, spatially dependent dissipation has been engineered using near-resonant optical fields and state-dependent atom loss, providing a direct route to implementing non-Hermitian single-particle Hamiltonians~\cite{Labouvie2016Bistability,Takasu2020,Li2019}. In particular, Takasu \textit{et al.} experimentally realized a $\mathcal{PT}$-symmetric many-body system using a Bose-Einstein condensate of $^{174}$Yb atoms in an optical superlattice, where controlled one-body loss was used to generate the dissipative component of the effective Hamiltonian \cite{Takasu2020}. Similarly, $\mathcal{PT}$-symmetry-breaking transitions have been observed in ultracold $^{6}$Li atoms through engineered state-dependent and time-dependent atom loss \cite{Li2019}. These experiments suggest that realizing spatially controlled dissipative potentials in atomtronic circuits is highly possible in the near future~\cite{ppvl-5lmg}.

In all results presented in this manuscript, the cavity field is turned on just after the barrier motion, following conventional phase contrast imaging \cite{levy2007acdc} or the absorption imaging-based methods ~\cite{RyuPRL2013,ryu2020quantum,bernhart2024observation,KSgan2025josephson2D,singh2026weaklinktunnelingcrossoveratomic}, in which the condensate image is taken when the barrier motion stops. By following the homodyne detection technique of cavity optomechanics \cite{KumarPRL2021}, we obtain the power spectrum of the phase quadrature of the cavity output field ($S(\omega)=\left|\mathrm{Im}\left[\alpha_{out}(\omega)\right]\right|^{2}$) after a simulation time $t_{\mathrm{cav}}$ with the optical barriers kept constant. The subsequent analysis of peak locations in the cavity output spectrum enables non-destructive detection of the Josephson oscillation frequencies in the atomtronic circuit~\cite{pradhan2025fractional, Pradhan2026diode}.

% Note that for the present ring geometry, 

\section{Josephson dynamics with $\mathcal{PT}$-symmetric potential}
\label{sec:PT}

%\subsection{Complex-potential-induced dynamics and cavity response}

We first examine the effect of the complex $\mathcal{PT}$-symmetric potential on the condensate dynamics of the atomtronic SQUID for static optical barriers, $f_{\mathrm{bar}}=0$. The corresponding condensate dynamics and the phase quadrature of the cavity output spectrum for three different values of $\Gamma_{\mathrm{local}}$ are shown in Fig.\ref{fig:fig1}. For $\Gamma_{\mathrm{local}}=0$, the system is effectively Hermitian and the condensate dynamics are governed by the optical barriers and the intrinsic Josephson coupling of the atomtronic SQUID. As shown in Fig.\ref{fig:fig1}(a)-(b), the condensate density exhibits only weak temporal modulation. The corresponding population imbalance in Fig.\ref{fig:fig1}(b) is likewise dominated by a small-amplitude Josephson dynamics, with no additional pronounced oscillatory component associated with gain and loss. This dynamical behaviour provides the reference configuration against which the effect of the complex PT-symmetric potential can be identified.

When a finite $\mathcal{PT}$-symmetric potential is introduced ($\Gamma_{\mathrm{local}}
\neq 0$), the dynamics change qualitatively. For $\Gamma_{\mathrm{local}}=0.5$, the gain and loss regions provide an additional channel for coherent population transfer, leading to a visible modulation of the condensate density in Fig.\ref{fig:fig1}(a,ii) and the associated population imbalance [Fig.\ref{fig:fig1}(b,ii)]. The amplitude of this oscillation increases with the non-Hermiticity (Fig.\ref{fig:fig1} (ii, iii)), demonstrating that the balanced gain-loss potential directly modifies the collective dynamics of the condensate. Since the gain and loss strengths are balanced, the resulting dynamics exhibit coherent oscillation between the source and drain regions rather than an overall amplification or depletion of the condensate. A similar balanced oscillation between the loss and gain regions in the presence of a PT-symmetric potential and the slowing down of oscillation for higher non-Hermiticity strength are also observed in the other systems \cite{Klaiman2008Visualization,ruter2010observation}. This reduction in frequency is a characteristic signature of the evolution of the $\mathcal{PT}$-symmetric eigenmodes toward their coalescence at the exceptional point, where the BdG eigenmodes become complex \cite{Cartarius2012Model}.
%---------------------------------------------------------------
 \begin{figure*}[!htp]
\begin{center}
\includegraphics[width= 1.0\linewidth]{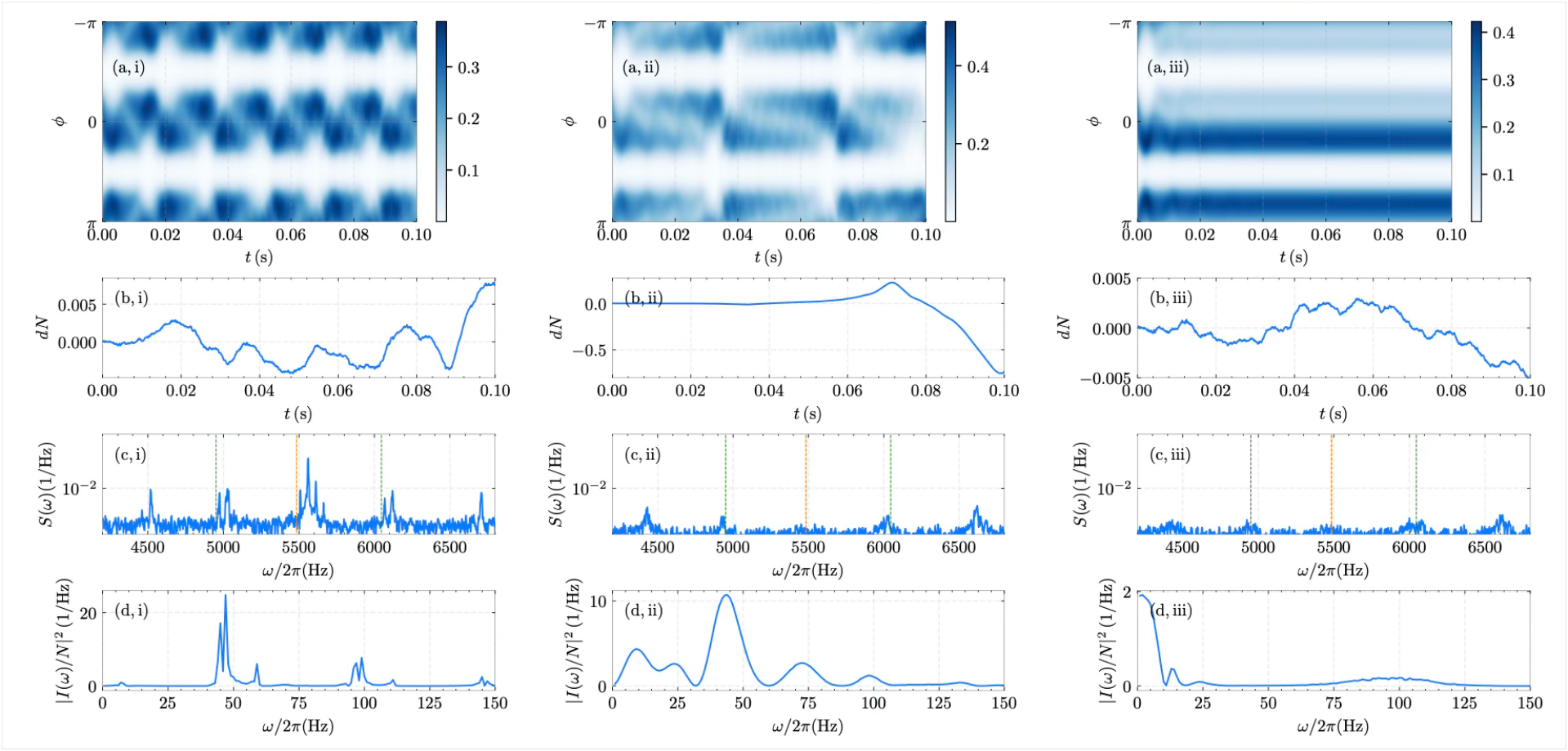} 
\end{center}
\vspace{-12pt}
\caption{Enhanced non-Hermitian dynamics of the atomtronic SQUID for $f_{\mathrm{bar}}=0$. Columns correspond to $\Gamma_{\mathrm{local}}=3.0$ (i) , $4.0$ (ii), and $5.0$ (iii). (a) Temporal evolution of the condensate density, (b) population imbalance, (c) power spectrum of the phase quadrature of the cavity output field, and (d) power spectrum of the population imbalance. In (c), the vertical dashed lines indicate the side-mode frequencies of uniform-density persistent-current states with $L_p=0$ (orange) and $L_p=\pm1$ (green) for $\ell=10$. The other set of parameters used is the same as those used in Fig.~\ref{fig:fig1}.}
\label{fig:fig2}
\end{figure*}
 %-------------------------------------------------------

The corresponding spectral response is shown in Fig.~\ref{fig:fig1}(c). The vertical dashed lines identify the side-mode frequencies associated with uniform-density persistent-current states ($\omega_{c,d}'=\left[\omega_{c,d}\left(\omega_{c,d}+4\tilde{g}N\right)\right]^{1/2}$, where $\omega_{c,d}(\Omega)=\omega_{\beta}\left(L_p\pm 2\ell\right)^{2}$ \cite{KumarPRL2021,pradhan2024cavity} ) with $L_p=0$ and $L_p=\pm1$ for $\ell=10$. In the absence of the complex potential, the cavity response is primarily determined by the side mode $L_p \pm 2\ell$ states for $L_p = 0$. As $\Gamma_{\mathrm{local}}$ is increased, additional peaks appear in the spectrum as a consequence of the $\mathcal{PT}$-induced density oscillations. For a high enough value of $\Gamma_{\mathrm{local}}$, the side modes split by capturing the condensate density oscillations  (see Fig.~\ref{fig:fig1}(c,iii)) and the corresponding magnitude of the split is the same as that of the dominant peak present in the power spectrum of population imbalance (see Fig.~\ref{fig:fig1}(d,iii)). From the Bogoliubov-de Gennes (BdG) analysis, we have verified that the PT-symmetric potential-induced density oscillation excites the collective modes (second eigenmode) of the condensate, whose values are shown in Fig.~\ref{fig:fig1}(d,iii), matching the magnitude of the split of the peaks in the cavity output spectrum. Thus, the cavity output provides a direct spectroscopic signature of the modification of the condensate dynamics by the presence of a PT-symmetric potential. In particular, the appearance and growth of the additional spectral feature with increasing $\Gamma_{\mathrm{local}}$ demonstrate that the PT-symmetric potential introduces a distinct dynamical channel in addition to the conventional plasma oscillation mode (first BdG eigen mode).

% The population-imbalance spectrum in Fig.~\ref{fig:fig1}(d) further resolves the characteristic frequencies of the system. The spectral features are compared with the eigenfrequencies obtained from the Bogoliubov-de Gennes (BdG) analysis, where both the optical barriers and the $\mathcal{PT}$-symmetric potential are included. The agreement between the dynamical spectral peaks and the BdG eigenvalues identifies the observed oscillations with collective excitation modes of the condensate. 

% Figure~\ref{fig:fig1} therefore demonstrates the onset of the $\mathcal{PT}$-induced dynamics as the non-Hermiticity is increased from zero. The simultaneous appearance of density modulation, population transfer, and an additional spectral response establishes the complex potential as an effective control parameter for the collective dynamics of the atomtronic SQUID. The stronger-non-Hermiticity regime is considered next in Fig.~\ref{fig:fig2}.
 
Having established the onset of the $\mathcal{PT}$-induced response for $\Gamma_{\mathrm{local}}\leq 2$ in Fig. \ref{fig:fig1}, we now consider the stronger non-Hermiticity regime. The corresponding condensate dynamics, population imbalance, and spectral response are shown in Fig.~\ref{fig:fig2}. These results demonstrate the progressive enhancement of the gain-loss-induced collective dynamics; however, for $\Gamma_{\mathrm{local}}>3.0$, the instability takes place (complex BdG modes), the oscillation frequency further slows down, and finally, the condensate density starts to amplify in only one region (see Fig.\ref{fig:fig2}(a,iii)). In this case, the absence of oscillations leads to only noise in the cavity output spectra, as shown in Fig. \ref{fig:fig2}(d,iii). Altogether, the decreasing condensate oscillation frequency and the absence of pronounced peaks in the cavity output spectrum signal the $\mathcal{PT}$-symmetry-breaking threshold in Figs.~\ref{fig:fig1} and \ref{fig:fig2}. In this way, the cavity spectroscopy provides a nondestructive probe of the condensate's evolution toward the exceptional point.

% The temporal evolution of the condensate density in Fig.\ref{fig:fig2}(a) shows a pronounced oscillatory response for all three values of $\Gamma_{\mathrm{local}}$. Compared with the weaker non-Hermiticity considered in Fig.\ref{fig:fig1}, the density modulation is substantially enhanced, indicating stronger population exchange between the gain and loss regions. At the same time, the oscillation period increases as $\Gamma_{\mathrm{local}}$ is increased from $3.0$ to $4.0$. Equivalently, the characteristic oscillation frequency decreases with increasing non-Hermiticity. 

% The corresponding population imbalance shown in Fig.~\ref{fig:fig2}(b) exhibits a pronounced periodic exchange between the gain and loss domains. The increasing amplitude of the imbalance demonstrates that the non-Hermitian potential increasingly dominates this component of the collective dynamics. In contrast to the conventional Josephson plasma oscillation associated with tunneling across the optical barriers, this higher-frequency collective response originates from the gain-loss structure of the $\mathcal{PT}$-symmetric potential. The two mechanisms can therefore be distinguished by their characteristic frequencies and spatial signatures.
%--------------------------------------------------------------------

\begin{figure*}[!htp]
\begin{center}
\includegraphics[width= 1.0\linewidth]{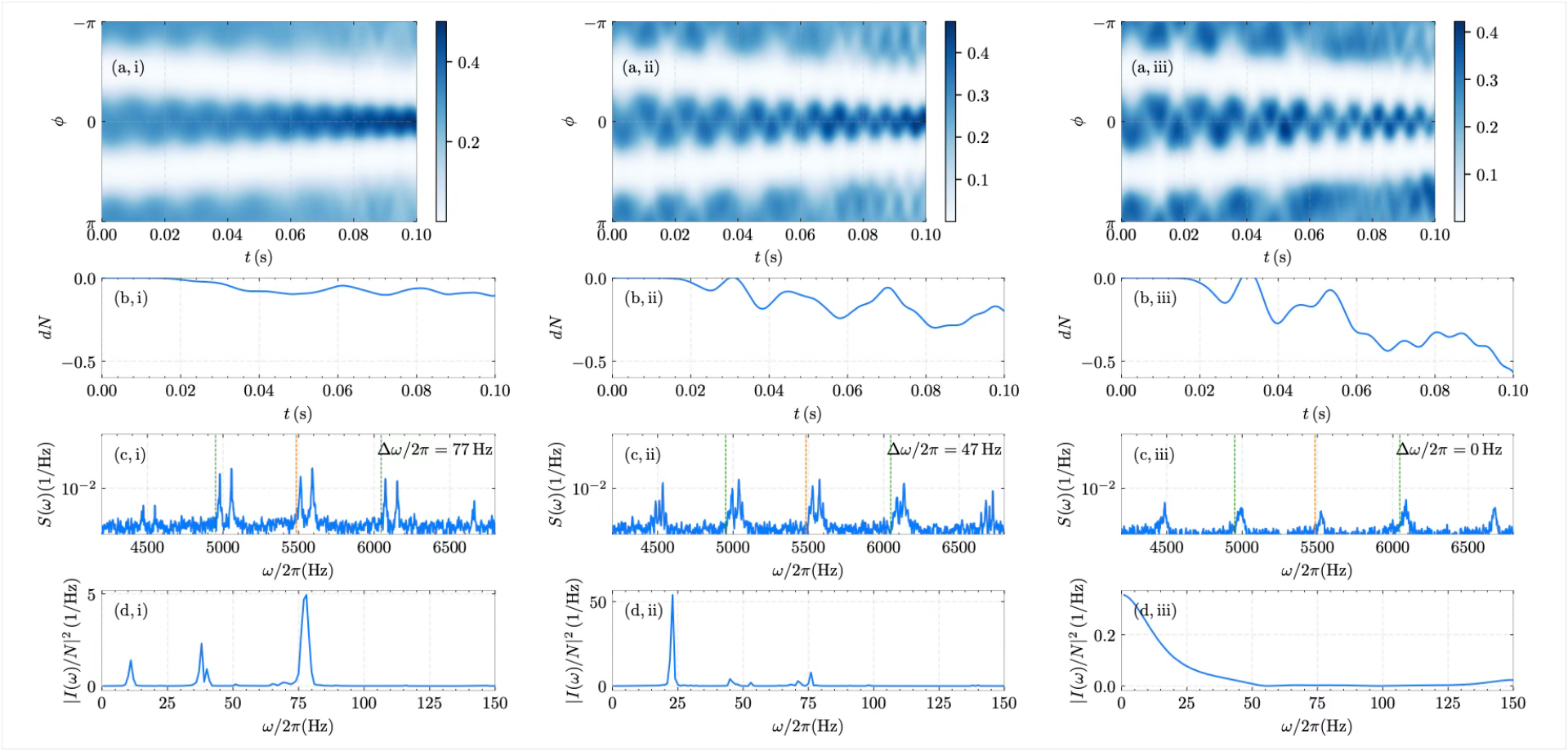} 

\end{center}
\vspace{-12pt}
\caption{Effect of the $\mathcal{PT}$-symmetric complex potential for $f_{\mathrm{bar}}=0.9~\mathrm{Hz}$ and $\Gamma_{\mathrm{local}}=0.5$ (i), $1.5$ (ii), and $2.0$ (iii). (a) Temporal evolution of the condensate density. (b) Temporal evolution of the population imbalance. (c) Power spectrum of the phase quadrature of the cavity output field. The vertical dashed lines indicate the side-mode frequencies corresponding to the uniform-density persistent-current states with $L_p=0$ (orange) and $L_p=\pm1$ (green) for $\ell=10$. (d) Power spectrum of the population imbalance. The other set of parameters used is the same as those used in Fig.~\ref{fig:fig1}.}
\label{fig:fig3}
\end{figure*}
 %-------------------------------------------------------

% Further evidence is provided by the power spectrum of the population imbalance in Fig.~\ref{fig:fig2}(d). The dominant spectral features are compared with the eigenfrequencies obtained from the BdG analysis incorporating both the optical barriers and the $\mathcal{PT}$-symmetric potential. The agreement between the spectral peaks and the calculated BdG eigenvalues confirms the collective-mode character of the observed oscillations. In particular, the spectral feature associated with the gain-loss-induced mode shifts toward lower frequency as $\Gamma_{\mathrm{local}}$ is increased. This softening of the mode provides a frequency-domain signature of the approach to the exceptional point.

% The same evolution is reflected in the cavity output spectrum shown in Fig.~\ref{fig:fig2}(c). Because the cavity field couples to the condensate density, the collective oscillations are imprinted on the phase quadrature of the output field. The vertical dashed lines again indicate the side-mode frequencies of the uniform-density persistent-current states, with the orange and green lines corresponding to $L_p=0$ and $L_p=\pm1$, respectively, for $\ell=10$. As the non-Hermiticity is increased, the cavity spectrum develops increasingly pronounced sideband structure associated with the $\mathcal{PT}$-induced density oscillations. The cavity therefore converts the collective matter-wave dynamics into a measurable optical signal without requiring direct destructive detection of the condensate.

We now consider the effect of spatially moving optical barriers in the presence of PT-symmetric potential. The dynamics for $f_{\mathrm{bar}}=0.9~\mathrm{Hz}$ and different values of $\Gamma_{\mathrm{local}}$ are shown in Fig.~\ref{fig:fig3}. For weak non-Hermiticity, the condensate dynamics are dominated by the ac Josephson effect for barrier velocity greater than a critical value~\cite{RyuPRL2013, levy2007acdc, KSgan2025josephson2D}, which results in a chemical potential difference in the density (Fig.~\ref{fig:fig3} (a,i)), particle oscillation between the two half rings (Fig.~\ref{fig:fig3} (b,i)), and the splitting of cavity output spectrum peaks with a magnitude the same as the Josephson oscillation frequency (Fig.~\ref{fig:fig3} (c,i) and (d,i)). When the non-Hermiticity is increased, the Josephson oscillation is dominated by the balanced loss-gain-induced particle oscillation, and chemical potential difference generation is suppressed or delayed; this is evident if we compare the density evolutions in Fig.~\ref{fig:fig3} (a,i), (a,ii), and (a,iii). Similarly, the population imbalance evolution also indicates that the critical population difference (minimum of $dN$) is higher for higher non-Hermiticity, as seen in Fig.~\ref{fig:fig3} (b,i), (b,ii), and (b,iii). The corresponding cavity output spectra also reveal the decrease in the chemical potential difference by the decreased magnitude of the splitting of the peaks for the same barrier velocities, as shown in Fig.~\ref{fig:fig3} (c,i), (c,ii), and (c,iii). Together, these results convey that the junction supports a higher critical barrier velocity or a higher critical current as the PT-symmetric potential strength increases, demonstrating a novel method for controlling the critical current in the atomtronic circuit. 

Another important aspect of the interplay between non-Hermiticity and translational motion of the barriers is that the non-zero imaginary eigenvalues appear at a smaller value of the loss-gain rate for $f_{\mathrm{bar}} = 0.9$ Hz (see Fig.~\ref{fig:fig3}) as compared to $f_{\mathrm{bar}} = 0.0$ Hz (see Fig.~\ref{fig:fig2}). This is also evident from the collective excitation spectrum of these two cases that are shown in Fig. \ref{fig:figA1} and Fig. \ref{fig:figA2}, respectively. Physically, the barrier displacement modifies the spatial overlap and localization of the collective modes, thereby changing their sensitivity to the non-Hermitian gain-loss potential. Thus, barrier motion brings the system closer to the $\mathcal{PT}$-symmetry-breaking transition, and due to this, the maximum allowed barrier velocity for dynamical stability tends to decrease as non-Hermiticity increases.

\begin{figure*}[!htp]
\begin{center}
\includegraphics[width= 1.0\linewidth]{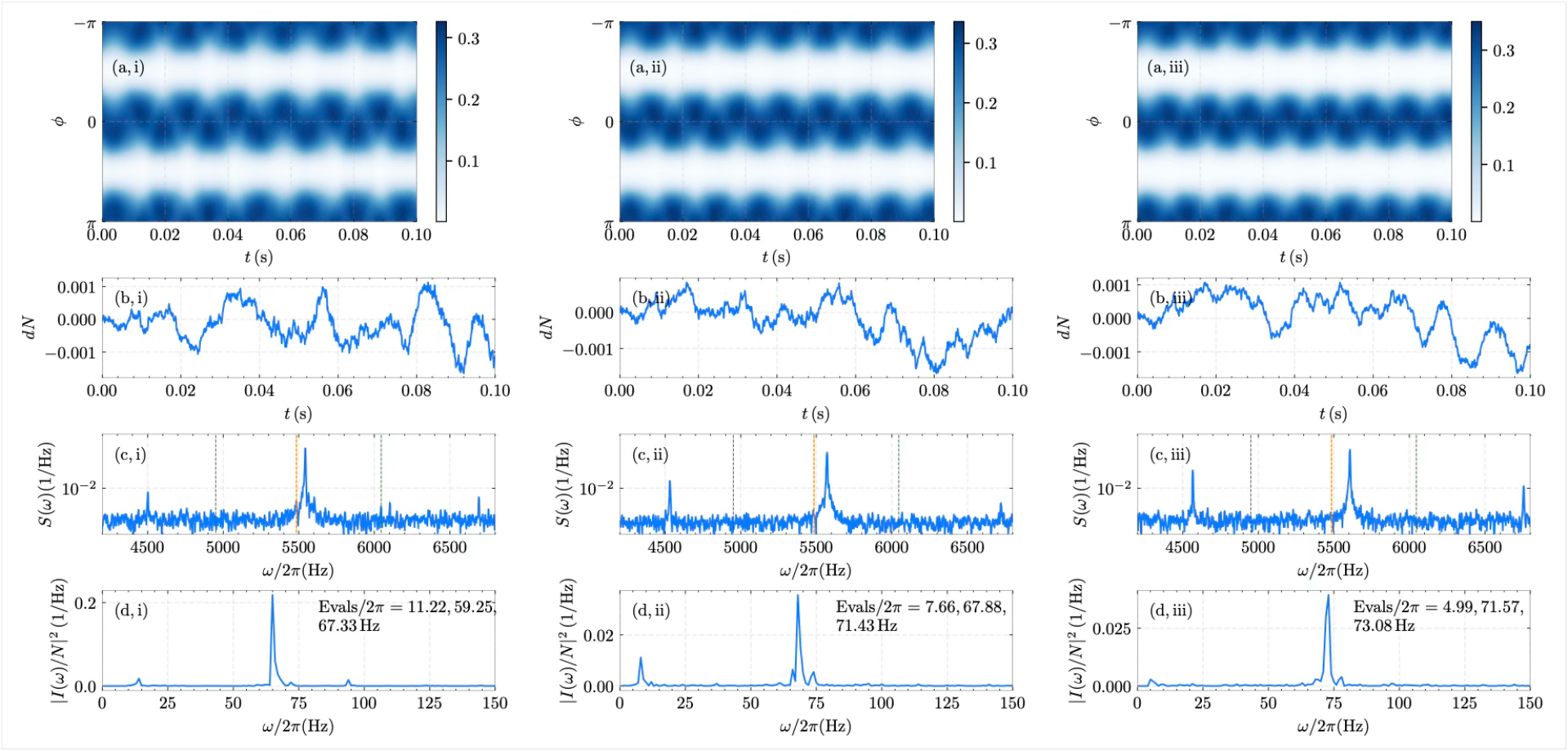} 

\end{center}
\vspace{-12pt}
\caption{Effects of barrier height (i) $V_0 = 12.0\,\mu_0$, (ii) $V_0 = 17.5\,\mu_0$, (iii) $V_0 = 24.0\,\mu_0$ for $\Gamma_{\mathrm{local}} = 1.0$ and $f_{\mathrm{bar}} = 0.0$ Hz. (a) Temporal evolution of condensate density, (b) Temporal evolution of population imbalance, (c) Power spectrum of phase quadrature of cavity output field (The vertical dashed lines indicate the side mode frequency corresponding to uniform density persistent current states with $L_p = 0 \, (\mathrm{orange})$ and $L_p = \pm 1 \, (\mathrm{green})$ for $\ell = 10$.), and (d) Power spectrum of population imbalance (Specified eigenvalues are obtained from BdG analysis incorporating optical barriers and PT-symmetric potential). The other parameters used are the same as those used in Fig.~\ref{fig:fig2}.} 
\label{fig:fig4}
\end{figure*}

To further establish the collective-mode character of the loss-gain-induced particle oscillations, we examine their dependence on the optical barrier height in Figure~\ref{fig:fig4} at fixed $\Gamma_{\mathrm{local}}$. The temporal evolution of the condensate density in Fig.~\ref{fig:fig4}(a) shows that increasing the barrier height modifies the characteristic time scales of the condensate dynamics. In particular, the oscillations associated with the $\mathcal{PT}$-symmetric potential become progressively faster as $V_0$ is increased. This behavior reflects the modification of the underlying collective-mode spectrum by the optical barriers. At the same time, increasing the barrier height suppresses tunneling between the two SQUID wells, thereby reducing the Josephson coupling and shifting the conventional Josephson plasma frequency to lower values.
%--------------------------------------------------------------------

 %-------------------------------------------------------
% The corresponding population imbalance in Fig.\ref{fig:fig4}(b) exhibits the same two characteristic dynamical scales identified in Figs.\ref{fig:fig1}--\ref{fig:fig3}. The lower-frequency component is associated with the Josephson plasma oscillation between the barrier-defined wells, whereas the faster oscillation originates from the gain-loss coupling generated by the $\mathcal{PT}$-symmetric potential. The opposite dependence of these two modes on the barrier height is particularly significant: increasing $V_0$ lowers the Josephson frequency while simultaneously increasing the frequency of the $\mathcal{PT}$-induced mode. Thus, the two modes respond differently to the same modification of the trapping landscape, providing a clear means of distinguishing their physical origins.

This behavior is confirmed quantitatively by the power spectrum of the population imbalance shown in Fig.~\ref{fig:fig4}(d). For each barrier height, the two dominant spectral features coincide with the corresponding BdG eigenfrequencies obtained by incorporating the optical barriers and the $\mathcal{PT}$-symmetric potential. The lowest BdG mode follows the Josephson plasma frequency and decreases systematically with increasing $V_0$, consistent with the reduction of the effective tunneling rate through a higher barrier. In contrast, the second BdG mode shifts to higher frequency with increasing barrier height and coincides with the frequency of the rapid population oscillations induced by the $\mathcal{PT}$ potential. The agreement between the time-domain oscillations and the BdG eigenvalues for all three barrier heights provides strong evidence that the observed dynamics originate from well-defined collective excitation modes. However, for the considered value of $\Gamma_{\mathrm{local}}$, the amplitude of loss-gain induced oscillation is not sufficient to result in the splitting of the side mode peaks in the cavity output spectrum, as seen from Fig.~\ref{fig:fig4}(c).

% The same mode structure is reflected in the cavity output spectrum [Fig.~\ref{fig:fig4}(c)]. Because the cavity phase quadrature is sensitive to fluctuations of the condensate density, the collective modes appear as spectral features in the optical response. The vertical dashed lines indicate the side-mode frequencies corresponding to the uniform-density persistent-current states with $L_p=0$ and $L_p=\pm1$ for $\ell=10$. The evolution of the cavity spectrum with $V_0$ follows the changes observed in the population-imbalance spectrum, demonstrating that the cavity provides a nondestructive probe of the barrier-dependent collective dynamics.

The distinct barrier-height dependence of the two modes also provides a useful physical interpretation of their origins. The Josephson plasma mode is controlled primarily by the tunneling energy between the two wells and therefore softens as the barrier is raised. The $\mathcal{PT}$-induced mode, in contrast, involves the spatial redistribution of density between the gain and loss regions and is consequently modified by the changing density profile and local mode structure generated by the optical barriers. Its increasing frequency with $V_0$ demonstrates that the non-Hermiticity-induced collective mode is intrinsically coupled to the barrier-defined geometry of the atomtronic SQUID rather than being an independent local oscillation.

\begin{figure}[!htp]
\begin{center}
\includegraphics[width= 1.0\linewidth]{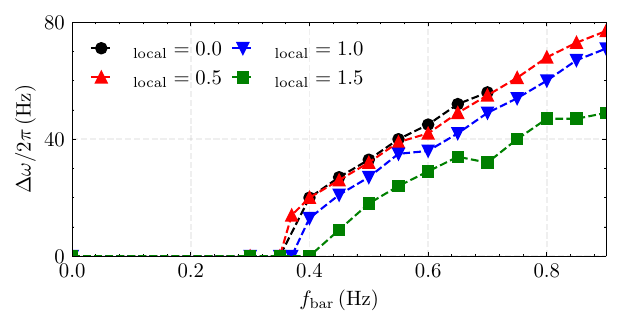}

\end{center}
\vspace{-12pt}
\caption{Effect of barrier motion on the cavity response. Magnitude of the peak splitting in the cavity output spectrum as a function of the barrier velocity $f_{\mathrm{bar}}$. The other parameters are the same as those used in Fig.~\ref{fig:fig2}. The emergence of a finite peak splitting signals the transition from the dc to the ac Josephson regime. Shapiro-like steps appear in the response in the absence of an externally applied ac drive, arising from the intrinsic $\mathcal{PT}$-induced oscillations. Here $f_m = 0$ Hz and $\phi_m = 0.0$ and the other set of parameters used is the same as those used in Fig.~\ref{fig:fig1}.}
\label{fig:PTdc-ac}
\end{figure}
 %-------------------------------------------------------
%--------------------------------------------------------------------

\begin{figure*}[!htp]
\begin{center}
\includegraphics[width= 1.0\linewidth]{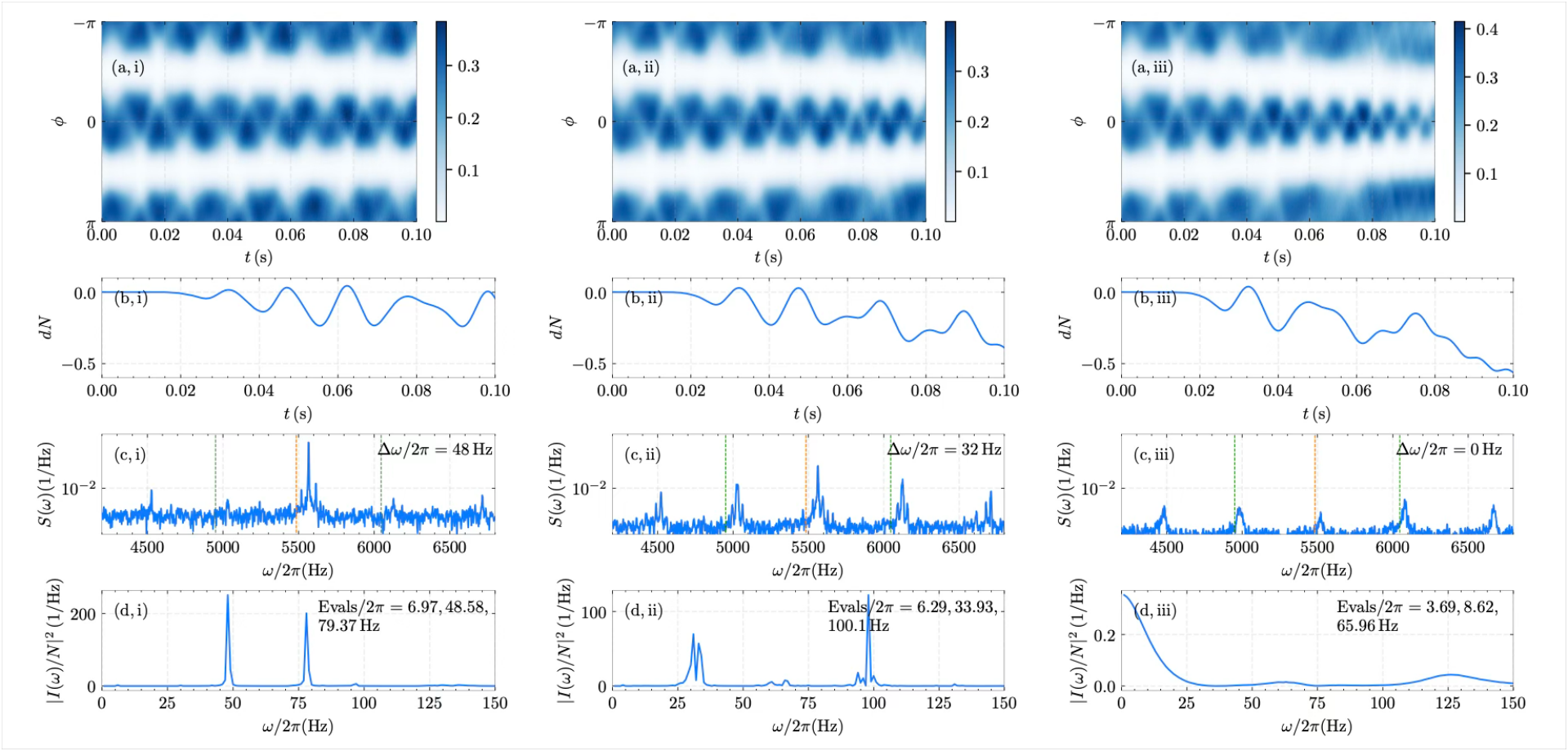} 
\end{center}
\vspace{-12pt}
\caption{Effects of barrier motion (i) $f_{\mathrm{bar}} = 0.25$ Hz , (ii) $f_{\mathrm{bar}} = 0.5$ Hz, and (iii) $f_{\mathrm{bar}} = 0.75$ Hz for $\Gamma_{\mathrm{local}} = 2.0$. (a) Temporal evolution of condensate density, (b) Temporal evolution of population imbalance, (c) Power spectrum of phase quadrature of cavity output field (The vertical dashed lines indicate the side mode frequency corresponding to uniform density persistent current states with $L_p = 0 \, (\mathrm{orange})$ and $L_p = \pm 1 \, (\mathrm{green})$ for $\ell = 10$.), and (d) Power spectrum of population imbalance (Specified eigenvalues are obtained from BdG analysis). The other parameters used are the same as those used in Fig.~\ref{fig:fig2}. The PT-symmetric oscillation frequency decreases with increasing $f_{\mathrm{bar}}$, and beyond $f_{\mathrm{bar}} = 0.75$, the eigenvalues become complex, and PT symmetry is broken. The other set of parameters used is the same as those used in Fig.~\ref{fig:fig1}. } 
\label{fig:fig5}
\end{figure*}
 %-------------------------------------------------------

% Overall, we find that the collective-mode spectrum provides a unified description of the dynamical behavior observed throughout the preceding sections. The BdG eigenfrequencies reproduce the characteristic oscillation frequencies of the condensate and population imbalance, while the corresponding modes are resolved nondestructively through the cavity output. Moreover, the systematic reduction of the exceptional-point threshold with increasing barrier modulation or AQUID asymmetry demonstrates that the onset of $\mathcal{PT}$-symmetry breaking can be controlled through the geometry of the atomtronic SQUID. These results establish the BdG spectrum as a direct link between the microscopic collective excitations, the observed Josephson and gain-loss dynamics, and the experimentally accessible cavity response.

\section{Current-Voltage Curve}
\label{sec:I-V}

%In this section, we demonstrate the effect of moving the two barriers towards each other around the ring. Previously, this barrier motion has been used for studying the DC-AC Josephson effect, critical current, current-phase relation, quantum interference of currents, and Shapiro steps. For a linear relative velocity between the condensate and the barriers and barrier velocity greater than a critical value, a chemical potential difference is generated in the system, giving rise to the ac Josephson effect. Below the critical velocity, the condensate remains in the dc regime and undergoes small-amplitude bounded phase oscillations, called the Josephson plasma oscillation. This critical velocity of the dc-ac Josephson transition is linked with the critical current of the junctions, which is being measured by plotting the chemical potential difference-barrier velocity curve (analog to the I-V curve). 

In this section, we investigate the current (barrier velocity) versus voltage (magnitude of peak splitting) curve to detect the dc-ac Josephson effect, Shapiro steps, and the quantum interference of currents in the presence of the PT-symmetric potential.

% Josephson dynamics generated by moving the two optical barriers toward each other around the ring. This protocol provides a controlled means of driving condensate flow through the junctions and has previously been used to study the dc–ac Josephson transition, critical current, current-phase relations, quantum interference of currents, and Shapiro steps. We drive the barriers linearly according to
% \begin{equation}
% \phi_{b_{1,2}}(t)=\pm(2\pi f_{\mathrm{bar}})t,
% \end{equation}
% up to $t_{\mathrm{bar}}$, while keeping the $\mathcal{PT}$-symmetric potential
% $V_{\mathrm{PT}}(\phi)=i\Gamma_{\mathrm{local}}\sin(\phi)$ fixed. For sufficiently small barrier velocities, the condensate remains in the dc Josephson regime, characterized by bounded relative-phase oscillations about a stable value and a vanishing time-averaged chemical-potential difference. These oscillations correspond to the Josephson plasma mode discussed above. As the barrier velocity exceeds a critical value $f_{\mathrm{c}}$, the phase develops a running component, and a finite chemical-potential difference $\Delta\mu=\mu_1-\mu_2$ emerges, signalling the onset of the ac Josephson regime. Thus, the $\Delta\mu$ versus $f_{\mathrm{bar}}$ characteristic provides the atomtronic analogue of an $I$-$V$ curve, with the barrier velocity representing the imposed current and $\Delta\mu$ the corresponding voltage, allowing the critical velocity and hence the critical current to be identified.

\subsection{Linear motion: dc-ac Josephson effect}
\label{subsec:linear_dc_ac}

In order to determine the critical barrier velocity of the dc-ac Josephson transition in the presence of the PT-symmetric potential, the current-voltage curve for different non-Hermiticity strengths $\Gamma_{\mathrm{local}}$ is shown in Fig.~\ref{fig:PTdc-ac}. The splitting remains zero below $f_{\mathrm{c}}$ and becomes finite above a critical value, which is related to the critical current of the junction \cite{KSgan2025josephson2D}. Thus, the cavity spectrum provides a direct spectroscopic signature of the dc-ac Josephson transition, similar to the previous works on the Hermitian setup~\cite{pradhan2025fractional, Pradhan2026diode}. However, in the present non-Hermitian setup, an important feature is that the critical barrier velocity increases with an increase in the non-Hermiticity strength, as discussed in the case of Fig.~\ref{fig:fig3}. Here, the balanced gain-loss potential modifies the condensate density and phase distribution around the ring through coherent population transfer between the gain and loss regions. Consequently, the phase response of the junctions to the imposed barrier motion is modified, allowing the condensate to sustain bounded phase dynamics (dc Josephson regime) over a larger range of barrier velocities. 

Another important observation is the emergence of plateaus in the current-voltage curve at higher values of $\Gamma_{\mathrm{local}}$. Similar plateaus have been observed in superconducting~\cite{shapiro1963, Vanneste1988} and atomtronic circuits~\cite{bernhart2024observation,delpace2024shapirofermi,singh2024, pradhan2025fractional} due to the microwave radiation and external barrier modulation, respectively. Similar plateaus are also observed in the atomtronic circuit when phonon modes are resonantly excited by the Josephson oscillations \cite{Bidasyuk2016phonon}. In the present case, we refer to these features as $\mathcal{PT}$-induced Shapiro-like steps because the synchronization is generated by the intrinsic oscillation associated with the $\mathcal{PT}$ potential rather than by an externally applied ac drive. As shown in Fig.~\ref{fig:fig5}, the Josephson oscillation frequency increases with $f_{\mathrm{bar}}$, whereas the frequency of the $\mathcal{PT}$-induced collective mode decreases. Their crossing therefore provides a natural resonance condition for phase locking. Near this resonance, the Josephson oscillation can synchronize with the $\mathcal{PT}$-induced oscillation, producing plateaus in the chemical-potential difference as the barrier velocity is varied. The increase in step width with $\Gamma_{\mathrm{local}}$ reflects the stronger coupling to the internal $\mathcal{PT}$ oscillation and the corresponding enhancement of the phase synchronization.

\subsection{Periodically Driven Josephson Dynamics}
\label{sec:Shapiro}

\begin{figure}[b]
\begin{center}
\includegraphics[width= 1.0\linewidth]{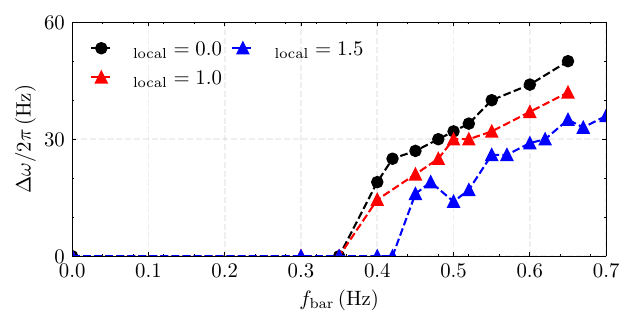}

\end{center}
\vspace{-12pt}
\caption{Effect of barrier position modulation on the cavity response.  Shapiro steps appear in the presence of an externally applied ac drive with $f_m = 30$ Hz and $\phi_m = 0.01$. The steps are more pronounced for higher PT-symmetric loss-gain strength. The other set of parameters used is the same as those used in Fig.~\ref{fig:fig1}.}
\label{fig:PTShapiro}
\end{figure}

We therefore consider a linear translation of the barriers with an additional periodic modulation (Eq.~\ref{Eq:bar_pos}), with modulation amplitude $\phi_{\mathrm{m}}$ and external modulation frequency $f_{\mathrm{m}}$. In contrast to the preceding case, where the oscillatory $\mathcal{PT}$ mode provides an intrinsic driving frequency, here the driving frequency $f_{\mathrm{m}}$ can be tuned independently. We consider the driving frequency to be less than the collective-mode frequency to ensure clear identification of the first Shapiro step. The corresponding current-voltage curve for different values of the non-Hermiticity strength is shown in Fig.~\ref{fig:PTShapiro}. Analogous to the increasing width of the Shapiro steps for lower barrier heights in other atomtronic circuits~\cite{bernhart2024observation}, the width of the Shapiro steps (plateaus at $\Delta \omega = 30$ Hz) increases for higher non-Hermiticity strength due to the increased coupling between the two half-rings facilitated by the loss-gain-induced oscillations. 

%--------------------------------------------------------------------

 %-------------------------------------------------------
 
%\subsection{Linear motion + rotation: Interference of currents and Shapiro steps}
\subsection{Combined translation and rotation of barrier positions: quantum interference, Shapiro-like steps, and diode effect}
\label{subsec:linear_rotation}

\begin{figure}[b]
\begin{center}
\includegraphics[width= 1.0\linewidth]{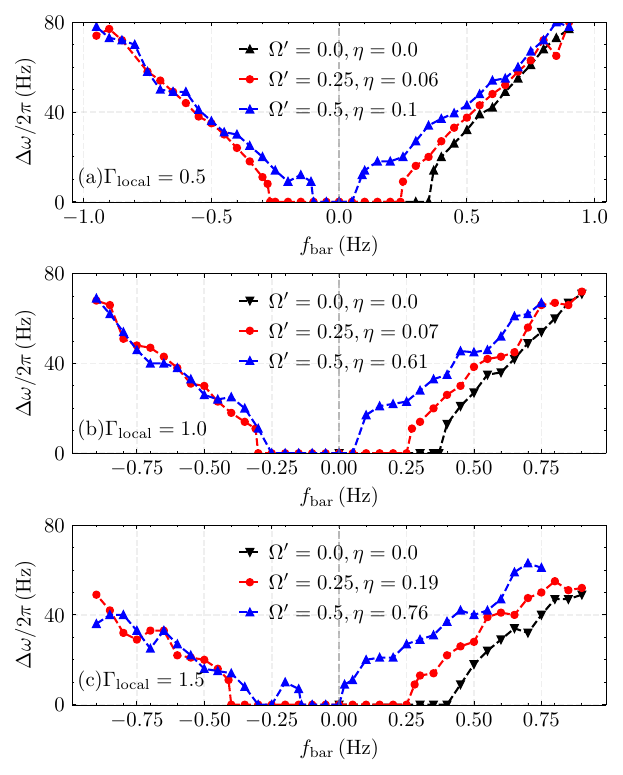}

\end{center}
\vspace{-12pt}
\caption{Effect of barrier motion in the rotating frame. Magnitude of the peak splitting in the cavity output spectrum, $\Delta\omega/2\pi$, as a function of the barrier velocity, $f_{\mathrm{bar}}$, for different values of the local decay rate: $\Gamma_{\mathrm{local}}=0.5$ (a), $1.0$ (b), and $1.5$ (c). All other parameters are fixed to the values used in Fig.~\ref{fig:fig2}. The other set of parameters used is the same as those used in Fig.~\ref{fig:fig1}. }
\label{fig:fig7}
\end{figure}

We next investigate the effect of simultaneously translating and rotating the two optical barriers. In contrast to the purely linear motion considered above, the additional rotation introduces a synthetic rotational flux and modifies the phase accumulated by the condensate around the ring~\cite{ryu2020quantum, Kiehn2022}. The combined motion therefore provides an additional control over the Josephson dynamics and allows us to investigate the interplay between quantum interference, non-Hermitian population transfer, and direction-dependent transport. Figure~\ref{fig:fig7} shows the current-voltage curve for different non-Hermiticity strengths $\Gamma_{\mathrm{local}}$ and rotation rate $\Omega'$. 

% Several characteristic features emerge from these curves, including oscillations of the critical velocity with rotation, self-induced Shapiro-like steps, and a pronounced nonreciprocal Josephson response.

% As the barriers are rotated around the ring, the condensate acquires an additional phase associated with the rotational motion, thereby changing the relative phase accumulated along the two possible current paths. The two junction currents consequently interfere constructively or destructively depending on $\Omega'$.

The oscillatory dependence of the critical velocity on the rotation rate is a direct manifestation of quantum interference of the currents circulating through the two junctions~\cite{ryu2020quantum, pradhan2025fractional}. In particular, the critical velocity reaches a minimum near the half-integer rotation rate $\Omega'=0.5$, corresponding to maximal destructive interference of the two circulating current contributions. This behavior is analogous to the flux dependence of the critical current in a conventional SQUID \cite{SQUIDbook} and demonstrates that the rotational motion provides a controllable means of tuning the effective Josephson coupling.
%--------------------------------------------------------------------

 %-------------------------------------------------------

The current-voltage curves exhibit additional step-like features that become increasingly pronounced as the rotation rate increases (see Fig.~\ref{fig:fig7}). These additional steps originate from the rotational motion of the barrier, which modifies the phase accumulated across the two junctions. Hence, the phases of the two junctions evolve differently with time and give rise to a different net voltage~\cite{Vanneste1988}. The average of the two voltage contributions from the two junctions results in fractional Shapiro steps and manifests as additional plateaus in the current-voltage curve~\cite{pradhan2025fractional}. In the current non-Hermitian setup, the combined effects of barrier rotation and PT-symmetric oscillation lead to the additional Shapiro steps in Fig.~\ref{fig:fig7}. The height and width of the steps further change by following the resonance between the Josephson oscillation and loss-gain-induced particle oscillation, as discussed in Fig \ref{fig:PTShapiro}.

% These features originate from synchronization between the Josephson oscillations driven by the translating barriers and the intrinsic density oscillations generated by the $\mathcal{PT}$-symmetric gain-loss potential. The latter therefore acts as an internal periodic drive for the Josephson dynamics, producing self-induced Shapiro-like steps even in the absence of an externally applied ac drive. The height and width of the steps increase with $\Omega'$, indicating that rotation enhances the coupling between the circulating Josephson current and the $\mathcal{PT}$-induced oscillatory mode. 

% Moreover, the rotational motion modifies the phase accumulated across the two junctions, allowing additional commensurability conditions between the Josephson and $\mathcal{PT}$ frequencies. As a result, multiple steps can appear at fractional positions as the rotation rate is varied. These fractional structures reflect higher-order synchronization processes and provide a characteristic signature of the coupled rotational and non-Hermitian dynamics.

A further consequence of the simultaneous translation and rotation is the emergence of the diode effect. In the present geometry, the two junctions are located in regions experiencing different parts of the gain-loss potential. Consequently, the phase jumps across the two junctions are inequivalent \cite{ryu2020quantum}. The combined action of rotation and translation consequently breaks the relevant time-reversal and inversion symmetries, leading to different critical velocities for opposite directions of barrier motion. This nonreciprocity is the hallmark of the Josephson diode effect.

% In the presence of rotation, the corresponding phase differences can be represented schematically as
% \begin{equation}
% \Phi_1=\Phi_0+\pi\Omega',
% \qquad
% \Phi_2=\Phi_0'-\pi\Omega',
% \end{equation}
% where the opposite signs of the rotational contribution reflect the opposite orientations of the two junctions around the ring. 

% The gain-loss potential therefore makes the two junctions dynamically inequivalent, while the rotation further modifies their phase response. When the barriers are simultaneously translated toward each other, the direction of motion determines the evolution of these phase differences. 

To quantify the directional asymmetry, we define the diode efficiency as
\begin{equation}
\eta=
\frac{|f_{c_+}|-|f_{c_-}|}
{|f_{c_+}|+|f_{c_-}|},
\end{equation}
where $f_{c_+}$ and $f_{c_-}$ denote the critical barrier velocities for motion in the positive and negative directions, respectively. The diode efficiency increases with $\Gamma_{\mathrm{local}}$, showing that non-Hermitian gain-loss coupling enhances the directional asymmetry of the Josephson response. For a fixed gain-loss strength, $\eta$ peaks near the half-integer rotation rate, $\Omega'=0.5$, where destructive interference between the circulating currents is strongest. Thus, the same rotational phase that suppresses the critical current also enhances the junction response's nonreciprocity. The simultaneous presence of rotation and balanced gain-loss coupling therefore provides a mechanism for converting the interference of persistent currents into a direction-dependent Josephson transport response.

Overall, we find that the combined linear and rotational motion of the barriers gives rise to a rich non-Hermitian Josephson response. The oscillation of the critical velocity with $\Omega'$ reveals quantum interference of the currents, while the appearance of self-induced Shapiro-like steps reflects synchronization between the barrier-driven Josephson mode and the intrinsic $\mathcal{PT}$-induced collective oscillation. At the same time, the inequivalent phase evolution of the two gain- and loss-associated junctions produces a directional asymmetry in the critical velocities, realising an atomtronic diode. These phenomena can be independently controlled via the barrier velocity, rotation rate, and gain-loss strength, demonstrating that the atomtronic SQUID provides a versatile platform for manipulating quantum interference and nonreciprocal transport in a non-Hermitian many-body system.

% The results above show that the intrinsic oscillatory mode generated by the $\mathcal{PT}$-symmetric gain-loss potential can act as an internal drive for the Josephson dynamics, giving rise to self-induced Shapiro-like structures. A natural question is how these synchronization effects are modified when an external periodic drive is applied directly to the barrier motion. Such a drive provides an independent control frequency and enables us to distinguish resonant synchronization with the intrinsic $\mathcal{PT}$ mode from off-resonant driving.

% \newpage
\section{Nonreciprocal Josephson dynamics in a tilted optical lattice}
\label{sec:titltedLattice}

\begin{figure}[b]
\begin{center}
\includegraphics[width= 1.0\linewidth]{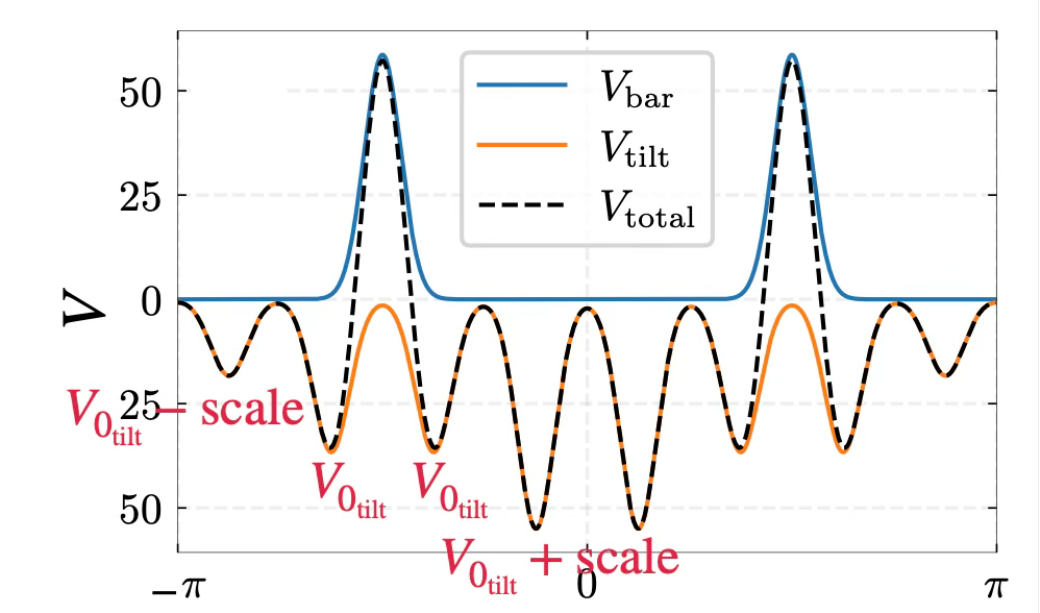}

\end{center}
\vspace{-12pt}
\caption{Tilted optical lattice: Potential energy profile for $8$ wells, where four of them near the two symmetrically placed optical barriers have the same depth $V_{0_{\mathrm{tilt}}} = 2.0\, \mu_0$, and the other four wells, having relatively more ($V_{0_{\mathrm{tilt}}} + \mathrm{scale}$) and less ($V_{0_{\mathrm{tilt}}} - \mathrm{scale}$) depth, act as the sink and source of the PT-symmetric potential. Here $R = 12 \mu$m, $V_0=4\mu_0$, $\sigma=1.4\zeta$, $\sigma_{\mathrm{tilt}} = 1.4 \zeta$ and $\mathrm{Scale} = 0.5 \mu_0$.} 
\label{fig:tiltedwell}
\end{figure}
 %-------------------------------------------------------
 
In the preceding sections, we considered the dynamics of a ring-shaped condensate in the presence of two moving optical barriers and a spatially engineered complex $\mathcal{PT}$-symmetric potential. We now consider an alternative, fully Hermitian realization in which a tilted optical lattice, together with the two Josephson barriers, provides the required spatial inversion-symmetry breaking and energy bias between different regions of the condensate~\cite{Kreibich2013Hermitian,Kreibich2014Realizing,Kreibich2016Tilted}. Unlike the complex gain-loss potential, the tilted lattice is described entirely by a conservative real potential and can therefore provide an experimentally accessible route for realizing biased Josephson transport without explicitly introducing gain or loss. The resulting energy imbalance establishes a preferred direction for particle transfer between the source and sink regions through the two Josephson junctions, allowing us to examine whether the characteristic non-reciprocal transport associated with the $\mathcal{PT}$-symmetric configuration can be reproduced in a Hermitian setting.
%--------------------------------------------------------------------
%--------------------------------------------------------------------

\begin{figure}[!htp]
\begin{center}
\includegraphics[width= 0.8\linewidth]{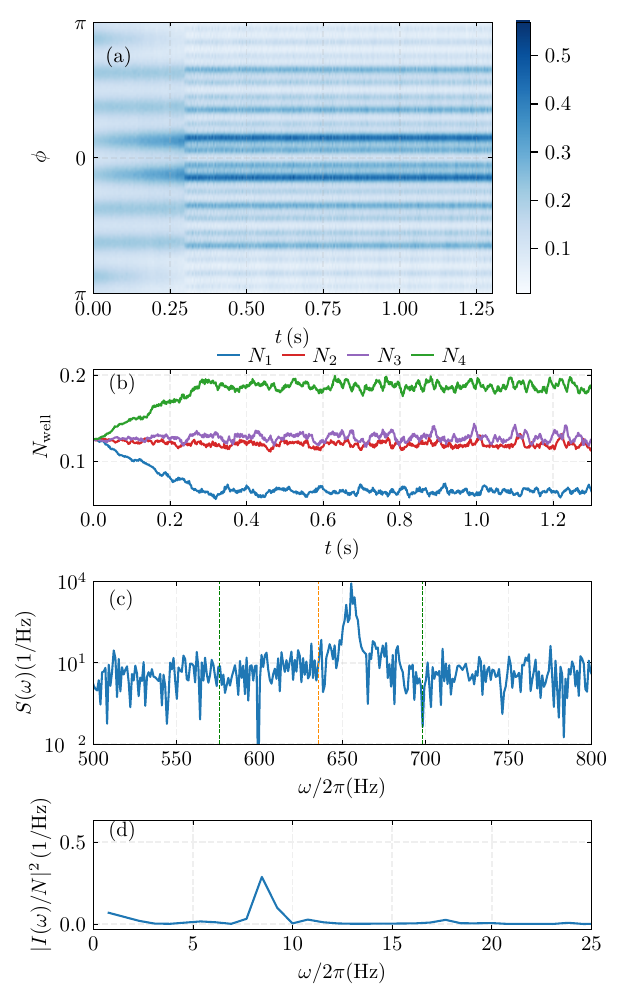}

\end{center}
\vspace{-12pt}
\caption{Creation and characterization of the source-sink configuration. (a) Temporal evolution of the condensate density during the linear ramp of the tilted-well depths from $V_{0_{\mathrm{tilt}}}$ to $V_{0_{\mathrm{tilt}}}\pm\mathrm{scale}$ over $t_{\mathrm{ramp}}=0.3$ s. After the ramp, the well depths are held fixed, and the cavity field is switched on for detection. (b) Corresponding populations of the four wells in the region $-\pi<\phi<0$. The source (sink) population decreases (increases) during the ramp, while the populations of the two symmetric wells remain nearly constant, establishing the desired source-sink configuration. (c) Phase-quadrature power spectrum of the cavity output, showing the dominant mode near $650\times2\pi$ Hz associated with the initial persistent-current state $L_p=0$. (d) Particle-current spectrum, $I(\omega)=\mathrm{FFT}[I(t)]$, with $I(t)=\partial n/\partial t$ and $n=(N_1-N_2)/(N_1+N_2)$, where $N_{1,2}$ are the populations of the two half-rings, separated by the symmetric positions of the optical barriers $\pm \pi/2$. However, the optical-barrier height is set to zero to isolate the effect of the tilted potential.  The set of parameters corresponding to the condensate is the same as used in Fig. ~\ref{fig:fig9} and the cavity parameters are the same as used in Fig.~\ref{fig:fig1}.} 
\label{fig:fig9}
\end{figure}
 %-------------------------------------------------------
To implement this configuration, we introduce a set of localized optical wells in addition to the two Josephson barriers. The tilted-well potential is modeled as a sum of Gaussian wells,
\begin{equation}
V_{{\mathrm{tilt}}}(\phi,\tau)=
-\sum_{j=1}^{N_{\mathrm{wells}}}
V_{{0_\mathrm{tilt}},j}(\tau)
\exp\left[
-2\left(
\frac{\phi-\phi_{b,j}}{\sigma_{\mathrm{tilt}}}
\right)^2
\right],
\label{Eq:Vtilt}
\end{equation}
where $V_{0,\mathrm{tilt},j}$ denotes the depth of the $j$th well, $\phi_{b,j}$ specifies its angular position along the ring, and $\sigma_{\mathrm{tilt}}$ determines the width of the wells. The positions and widths of all wells are kept fixed throughout the simulations, while their depths are varied to generate the desired spatial energy bias. The total trapping potential is obtained by combining $V_{\mathrm{tilt}}$ with the two moving optical-barrier potentials discussed previously.
%--------------------------------------------------------------------

\begin{figure*}[!htp]
\begin{center}
\includegraphics[width= 0.8\linewidth]{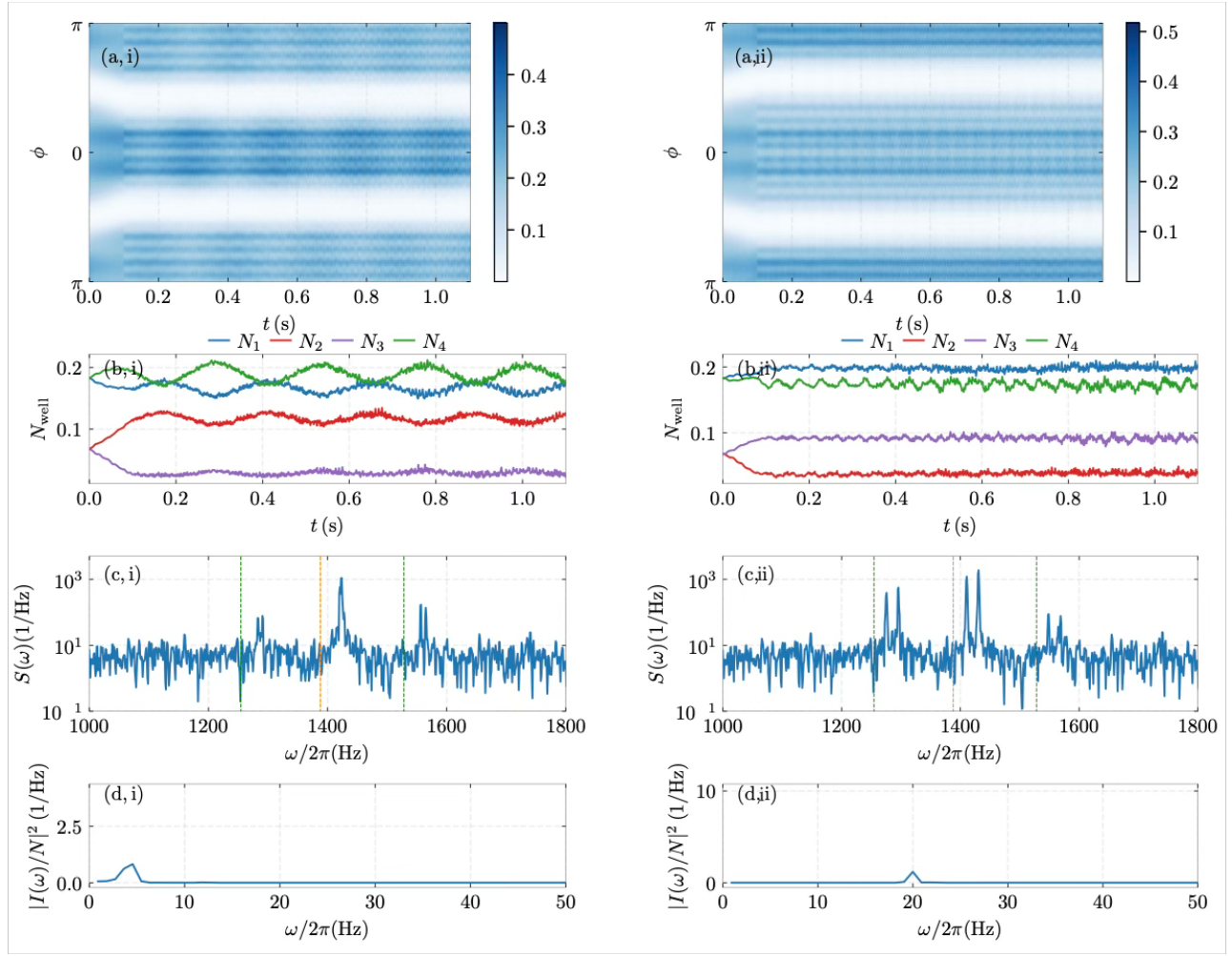} 

\end{center}
\vspace{-12pt}
\caption{Directional Josephson dynamics for positive and negative barrier velocities, $f_{\mathrm{bar}}=+0.35$ (i) and $f_{\mathrm{bar}}=-0.35$ (ii) with $V_{0_{\mathrm{tilt}}}=1.0\, \mu_0$, $\mathrm{scale} = 0.4\, \mu_0$, and $t_{\mathrm{ramp}}=0.1$ s. (a) Temporal evolution of the condensate density in the combined tilted-well and optical-barrier potentials. The barriers are driven toward each other with velocity $\pm f_{\mathrm{bar}}$ up to $t_{\mathrm{bar}}$, while the tilted-well depths are ramped to maintain the spatial bias. (b) Corresponding populations of the four wells in $-\pi<\phi<0$, showing distinct oscillation frequencies for the two transport directions. (c) Phase-quadrature power spectra of the cavity output. For positive bias, the $L_p=0$ peak remains unsplit, while the $L_p=1$ side peaks exhibit weak plasma-mode splitting, indicating a dc Josephson regime. For negative bias, all peaks split, with the additional splitting corresponding to the Josephson frequency, indicating the onset of the ac regime. (d) Particle-current spectra, showing the plasma oscillation for positive bias and an additional Josephson-frequency component for negative bias. The set of parameters used here are the same as used in Fig. ~\ref{fig:fig9}.}
\label{fig:fig10}
\end{figure*}
 %-------------------------------------------------------
We consider four auxiliary wells arranged symmetrically about the two Josephson barriers. The two central wells have equal depths, whereas the outer wells are biased such that one side of the barrier corresponds to a higher-energy region and the opposite side to a lower-energy region. This configuration effectively creates source and sink regions connected through the Josephson junctions. The resulting energy imbalance serves as the bias introduced by the gain and loss regions in the $\mathcal{PT}$-symmetric configuration, while the dynamics are still governed by a Hermitian Hamiltonian \cite{Kreibich2013Hermitian,Kreibich2014Realizing,Kreibich2016Tilted}. By reversing the tilt direction, the sign of the energy bias can be reversed, allowing the transport response for opposite current directions to be compared directly.

To minimize transient density oscillations associated with the sudden introduction of the tilted potential, the well depths are ramped linearly in time from their initial common value. Specifically, the wells are initialized with the same depth $V_0$, after which the depths are gradually changed to the biased values
\begin{equation}
\left(
V_{0_{\mathrm{tilt}}}-\mathrm{scale},
V_{0_{\mathrm{tilt}}},
V_{0_{\mathrm{tilt}}},
V_{0_{\mathrm{tilt}}}+\mathrm{scale}
\right),
\end{equation}
for one orientation of the tilt, with the corresponding mirror configuration used for the opposite orientation. Here, $\mathrm{Scale}$ controls the magnitude of the imposed energy bias. The linear ramp allows the condensate density to adjust smoothly to the modified trapping landscape and reduces unwanted excitations that could obscure the subsequent Josephson dynamics. The combined potential produced by the optical barriers and the tilted wells is shown in Fig.\ref{fig:tiltedwell}.

% After the ramp is completed, the system is allowed to evolve in the resulting stationary biased configuration before the barrier motion is applied.

%  The corresponding stationary condensate density is displayed in Fig.\ref{fig:tiltedwell}(b). The density remains localized around the wells while the energy offset between the higher- and lower-energy regions establishes a preferred direction for population transfer across the Josephson junctions. The tilted lattice therefore provides a controlled Hermitian analogue of the biased transport generated by the $\mathcal{PT}$-symmetric potential.

% We next use this configuration to investigate the response of the Josephson junctions to oppositely directed barrier motion. By comparing the critical barrier velocities for the two bias directions, we characterize the resulting non-reciprocal transport and determine the conditions under which a Josephson diode response emerges. This comparison allows us to distinguish effects arising from the spatial energy bias itself from those specifically associated with complex gain-loss coupling, and provides a direct test of how far the essential transport physics of the $\mathcal{PT}$-symmetric system can be reproduced using a purely Hermitian tilted lattice. 
%--------------------------------------------------------------------

\begin{figure*}[!htp]
\begin{center}
\includegraphics[width= 0.8\linewidth]{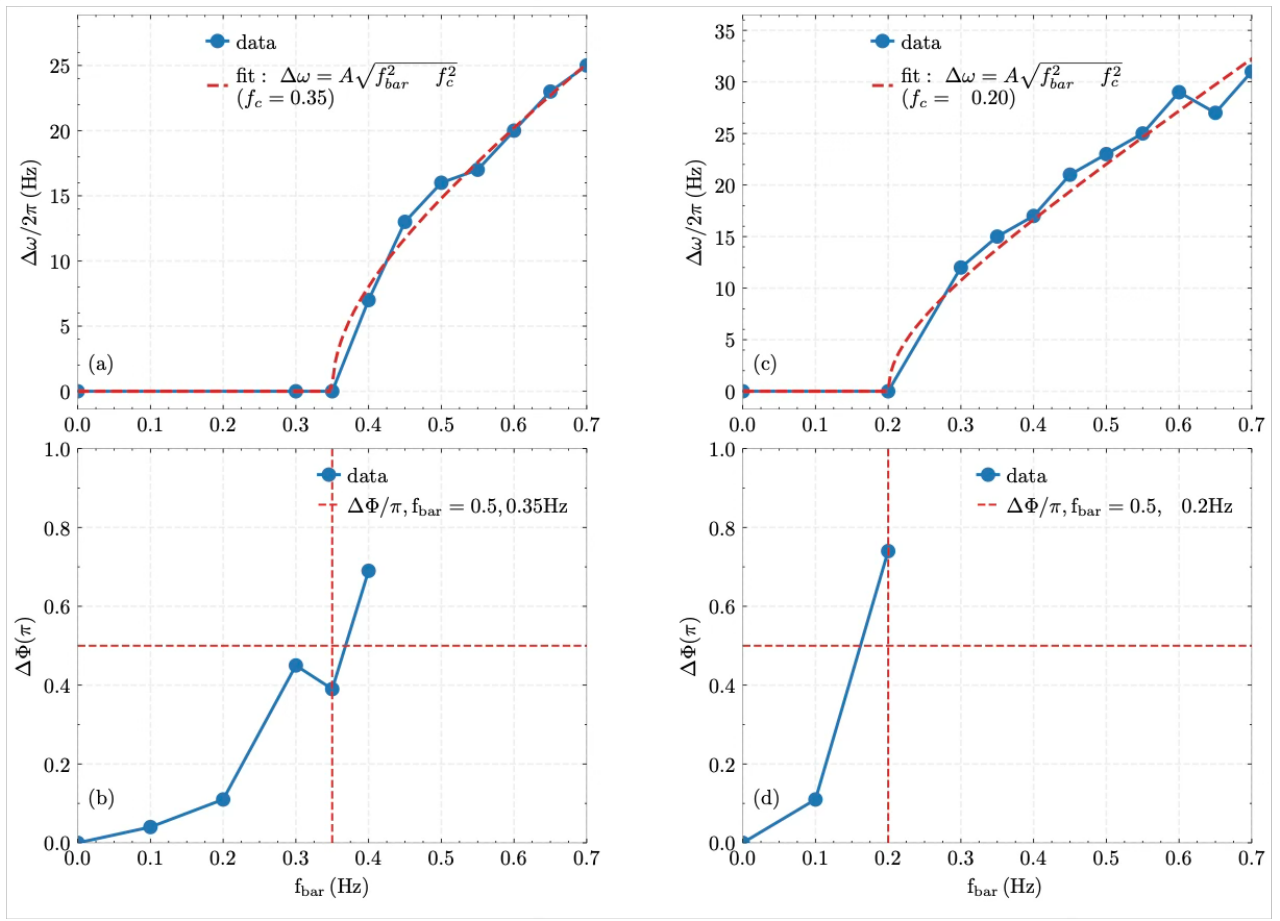} 

\end{center}
\vspace{-12pt}
\caption{Determination of the critical barrier velocity for positive bias, (a,b), and negative bias, (c,d). (a,c) Splitting $\Delta\omega$ of the side-mode peaks in the cavity-output spectrum as a function of barrier velocity $f_{\mathrm{bar}}$, providing the current-voltage characteristics. The critical velocity $f_c$ is extracted from $\Delta\omega=A\sqrt{f_{\mathrm{bar}}^2-f_c^2}$. (b,d) Phase jump $\Delta\Phi$ across the barrier after the barrier motion is stopped, showing the current-phase relation in the dc regime. The crossing of $|\Delta\Phi|=\pi/2$ identifies $f_c$, consistent with the Josephson relation $I=I_c\sin(\Delta\Phi)$.  Here $V_{0_{\mathrm{tilt}}}=1.0\, \mu_0$, $\mathrm{scale} = 0.2\, \mu_0$ and the set of parameters used here are the same as used in Fig. ~\ref{fig:fig9}.}
\label{fig:fig11}
\end{figure*}
 %-------------------------------------------------------
To establish the source and sink configuration, we first examine the temporal response of the condensate during the ramping of the tilted-well potential. Figure~\ref{fig:fig9}(a) shows the evolution of the condensate density as the depths of the four wells are varied linearly from their initial value $V_{0_{\mathrm{tilt}}}$ to $V_{0_{\mathrm{tilt}}}\pm\mathrm{Scale}$ over a ramp time $t_{\mathrm{ramp}}=0.3$ s. After the ramp is completed, the well depths are held fixed and the cavity field is switched on for the subsequent detection of the condensate dynamics. The corresponding population evolution of the four wells is shown in Fig.~\ref{fig:fig9}(b). During the ramp, the populations of the source and sink wells decrease and increase, respectively, while the populations of the two symmetric wells remain approximately unchanged. This behavior demonstrates that the tilted potential establishes a controlled population imbalance while maintaining approximately balanced particle transfer through the intermediate wells. In this sense, the tilted lattice provides a Hermitian realization of the source-sink configuration required for biased Josephson transport.
 
The corresponding cavity and matter-wave spectra provide complementary signatures of the prepared state. With the optical-barrier height set to zero, the phase-quadrature spectrum of the cavity output in Fig.\ref{fig:fig9}(c) exhibits a dominant peak near $650\times2\pi$ Hz, associated with the initial persistent-current state $L_p=0$. To isolate the dynamical response of the condensate generated by the tilted potential, we also calculate the particle-current spectrum $I(\omega)$, which gives a peak at the plasma oscillation frequency induced by the barrier motion, similar to the PT-symmetric potential case. Together, these results confirm that the tilted lattice generates the desired biased population transfer without requiring an explicitly complex potential.

% \begin{equation}
% I(\omega)=\mathrm{FFT}[I(t)],
% \qquad
% I(t)=\frac{dn}{dt},
% \end{equation}
% where
% \begin{equation}
% n=\frac{N_1-N_2}{N_1+N_2},
% \end{equation}
% and $N_{1,2}$ denote the populations of the two half-rings defined by the optical barriers. The resulting spectrum, shown in Fig.\ref{fig:fig9}(d), provides a direct measure of the population-transfer dynamics. 

We next investigate how this bias modifies the Josephson response by comparing opposite directions of barrier motion in Fig.~\ref{fig:fig10}. The optical barriers are translated toward one another across the ring, while the tilted-well depths are ramped to maintain the required spatial bias during the evolution. A pronounced directional asymmetry is observed in the condensate dynamics [Fig.~\ref{fig:fig10}(a,i) and (a,ii)]. In particular, the population oscillations of the four wells [Fig.~\ref{fig:fig10}(b,i) and (b,ii)] occur at different frequencies for opposite barrier velocities, demonstrating that the energy bias couples differently to the two directions of Josephson transport.
%--------------------------------------------------------------------

\begin{figure}[!htp]
\begin{center}
\includegraphics[width= 1.0\linewidth]{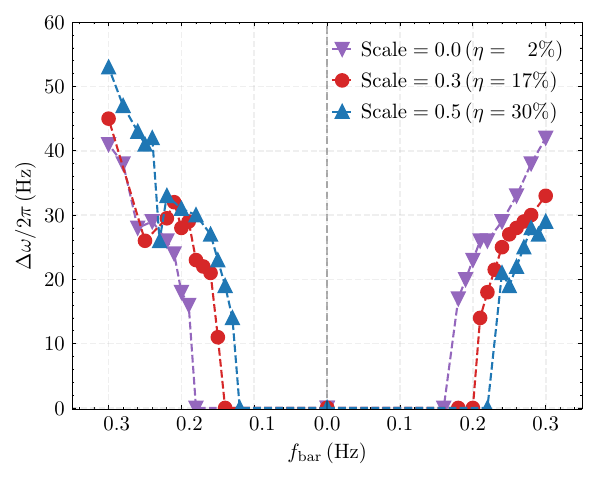} 
\end{center}
\vspace{-12pt}
\caption{Josephson diode efficiency. Magnitude of the side-mode splitting $\Delta\omega$ in the cavity-output spectrum as a function of positive and negative barrier velocities $f_{\mathrm{bar}}$. The directional critical velocities $f_{\mathrm{c}\pm}$ are used to quantify the diode efficiency, $\eta=(|I_{\mathrm{c}+}|-|I_{\mathrm{c}-}|)/(|I_{\mathrm{c}+}|+|I_{\mathrm{c}-}|)=(|f_{\mathrm{c}+}|-|f_{\mathrm{c}-}|)/(|f_{\mathrm{c}+}|+|f_{\mathrm{c}-}|)$. The diode efficiency increases by increasing the relative depth of the wells ($\mathrm{Scale}$), which breaks the spatial inversion symmetry. The nonzero diode efficiency for the symmetric potential profile ($\mathrm{Scale}=0.0$) arises from thermal fluctuations of the condensate. Here $R = 8\,\mu$m, $N = 7000$, $V_0 = 6.0\mu$, and $\sigma = 1.4 \zeta$. }
\label{fig:fig12}
\end{figure}
 %-------------------------------------------------------
 \begin{figure}[b]
\begin{center}
\includegraphics[width= 1.0\linewidth]{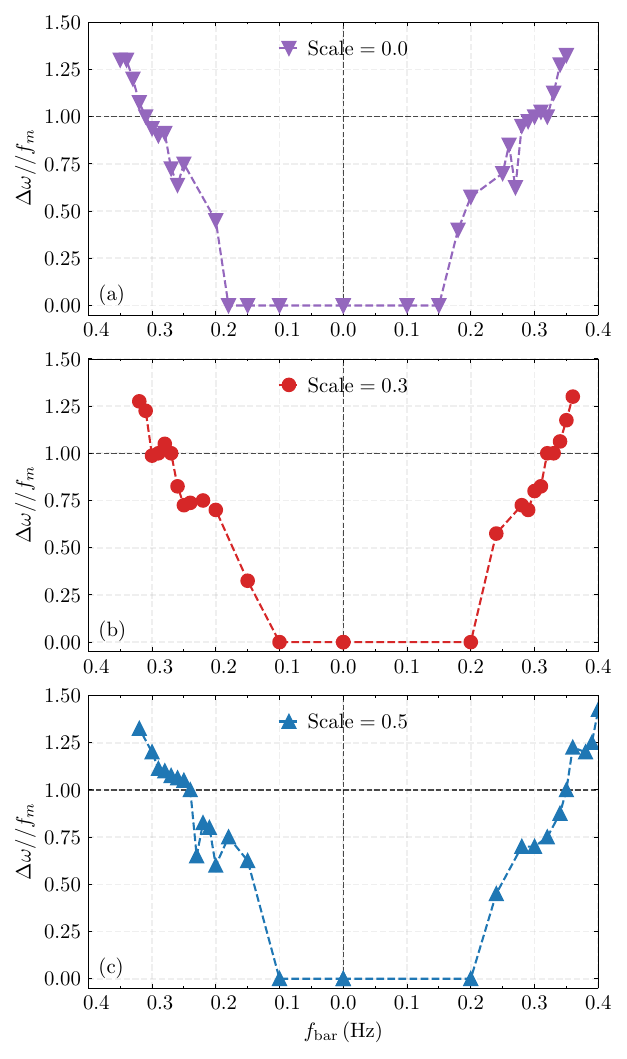}

\end{center}
\vspace{-12pt}
\caption{Shapiro steps for different tilt strengths: (a)Scale=0.0, (b) Scale=0.3, and (c) Scale=0.5. Magnitude of the side-mode splitting $\Delta\omega$ in the cavity-output spectrum as a function of positive and negative barrier velocities $f_{\mathrm{bar}}$ for different values of the relative depth of the wells $\mathrm{Scale}$. The resulting plateaus indicate phase locking between the Josephson oscillations of the particles and the external barrier modulation. The width of the steps becomes different for positive and negative biasing (asymmetric Shapiro steps) when $\mathrm{Scale} \neq 0.0$. The set of parameters used here is $f_m = 40$ Hz, $\phi_m = 0.01$, $R=8,\mu$m, $N=7000$, $V_0=6\mu_0$, and $\sigma=1.4\zeta$, and the other parameters are the same as those used in Fig.~\ref{fig:fig12}.}
\label{fig:fig13}
\end{figure}
 %-------------------------------------------------------

This directional dependence is also clearly visible in the cavity output spectrum (see Fig.~\ref{fig:fig10}(c,i) and (c,ii). For positive barrier motion, the central $L_p=0$ peak remains essentially unsplit, while the side peaks associated with the $L_p=1$ circulating-current state exhibit a small splitting due to the weak-amplitude plasma oscillation. The system therefore remains in the dc Josephson regime. In contrast, for negative barrier motion, all spectral peaks split, with the additional splitting corresponding to the Josephson oscillation frequency. This indicates that the system has entered the ac Josephson regime for the same magnitude of barrier velocity but in the opposite direction. The particle-current spectra in Fig.~\ref{fig:fig10}(d) provide an independent confirmation of the above-mentioned effect. In the positive-bias case, the plasma oscillation mode is dominant, whereas an additional Josephson-frequency component is dominant in the negative-bias case. Thus, the tilted lattice converts the spatial energy bias into a direction-dependent dynamical threshold, providing a direct mechanism for non-reciprocal Josephson transport.

To quantify this directional transition, we determine the critical barrier velocity for each bias direction from the cavity-output spectrum. As shown in the upper panels of Fig.~\ref{fig:fig11}, the splitting $\Delta\omega$ of the side-mode peaks increases above a well-defined threshold. We extract the critical velocity $f_c$ by fitting the numerical data to $$
\Delta\omega=
A\sqrt{f_{\mathrm{bar}}^2-f_c^2}$$, 
which is the characteristic square-root dependence expected near the onset of Josephson oscillations \cite{bernhart2024observation}, and $A$ is proportional to the resistance of the circuit. The lower panels of Fig.~\ref{fig:fig11} provide an independent characterization by measuring the current-phase relation of the circuit \cite{bernhart2024observation}. We plot the phase jump across the barrier after the barrier motion is stopped as a function of the barrier velocity $f_{\mathrm{bar}}$ and the transition occurs when the phase jump reaches $\pi/2$, consistent with the sinusoidal Josephson current-phase relation. 

% Here, the agreement between the spectroscopic and phase-based criteria provides a consistent determination of the critical velocity and establishes the cavity splitting as a useful nondestructive measure of the Josephson critical current. It's important to note that the above criteria for detecting critical velocities can also be applied to the PT-symmetric potential case.

By comparing the results in Fig.~\ref{fig:fig11} (i) and (ii), it's clear that due to the tilted wells, the critical velocities in the positive and negative directions differ by a non-zero amount, and this non-reciprocity is quantified through the diode efficiency, defined as
\begin{equation}
\eta=
\frac{|I_{c+}|-|I_{c-}|}
{|I_{c+}|+|I_{c-}|} =
\frac{|f_{c+}|-|f_{c-}|}
{|f_{c+}|+|f_{c-}|},
\end{equation}
where $I_{c\pm}$ ($f_{c\pm}$) denote the critical currents (critical barrier velocities) for the two directions of transport. 

The Josephson oscillation frequency vs positive and negative barrier velocities (voltage-current) curve and the critical velocities for higher tilt strength are shown in Figure~\ref{fig:fig12}. Importantly, the diode response here originates from the controlled spatial energy bias of the tilted lattice, demonstrating that directional Josephson transport does not require an explicitly non-Hermitian potential. Also, the non-reciprocity can be easily tuned in the experiments by increasing the tilt of the external bias potential.

% Figure~\ref{fig:fig12} shows the splitting $\Delta\omega$ as a function of the  barrier velocities. The unequal onset of the spectral splitting directly reflects the difference between the two critical velocities and hence provides a spectroscopic characterization of the Josephson diode effect. 

Finally, we investigate the system's response to periodic modulation of the barrier position. As shown in Fig.~\ref{fig:fig13}, the cavity-output peak splitting exhibits plateau-like features as the barrier velocity is varied, indicating synchronization between the externally imposed modulation and the intrinsic Josephson dynamics. These plateaus are the Shapiro steps as observed in the conventional Josephson junctions \cite{shapiro1963} and recently in atomtronic circuits \cite{bernhart2024observation, delpace2024shapirofermi, singh2024, pradhan2025fractional} , where an external periodic drive locks the Josephson oscillations to the drive frequency. 

% Here, we focus on a modulation frequency of $f_m=50$ Hz, for which clear integer Shapiro steps are observed; such steps were not obtained for the other modulation frequencies investigated.

The synchronization response depends strongly on the tilt strength of the optical lattice. For the untilted case ($\mathrm{Scale}=0$) as shown in Fig.~\ref{fig:fig13} (a), the steps are approximately symmetric with respect to the direction of barrier motion. In contrast, when a finite tilt is introduced [panels (b) and (c)], the widths and positions of the steps become different for positive and negative barrier velocities. This directional asymmetry reflects the diode effect, arising from the different transport thresholds and dynamical responses in the two directions. Increasing the relative depth difference between the wells therefore provides a means of tuning the degree of asymmetry and the robustness of the synchronized states. Thus, the tilted optical lattice provides a conservative mechanism for controlling the synchronization of Josephson dynamics with an external periodic drive, giving rise to direction-dependent, Shapiro-like phase locking. The results demonstrate that the spatial energy bias can be used to control not only the directional Josephson response but also the resulting non-reciprocal synchronization dynamics.

In a nutshell, we establish that a purely Hermitian tilted optical lattice can reproduce the essential biased-transport characteristics required for realising an atomtronic diode while providing direct access to the underlying population oscillation, phase evolution, and the cavity output spectrum. The agreement between the current-spectrum, cavity-output spectrum, and phase-based signatures further establishes the tilted lattice as a complementary realization of the biased Josephson system and provides an experimentally accessible route for studying non-reciprocal atomtronic transport without explicit gain and loss of atoms from the condensate.

 %--------------------------------------------------------------------

% The Josephson diode is realized as the signal for positive and negative biasing are different. 

   %--------------------------------------------------------------------
\section{Conclusion and outlook}
\label{sec:conclusion}

We have investigated non-Hermitian Josephson dynamics in an atomtronic SQUID formed by a ring Bose–Einstein condensate with two movable optical barriers. By introducing spatially balanced gain and loss, we establish an effective $\mathcal{PT}$-symmetric potential and show that non-Hermiticity provides a new mechanism for controlling both collective excitations and the matter-wave transport in atomtronic SQUID. In addition to the conventional Josephson plasma mode, the system supports a $\mathcal{PT}$-induced collective mode whose frequency decreases with increasing gain–loss strength and softens toward an exceptional point. The associated coherent population exchange leads to an additional splitting of the peaks in the cavity-output spectrum, providing a nondestructive spectroscopic signature of the non-Hermitian collective dynamics.

We further demonstrated that the interplay between $\mathcal{PT}$ symmetry and barrier motion leads to direction-dependent critical velocities, thereby yielding non-reciprocal Josephson transport. Simultaneous translation and rotation of the optical barriers provide additional control over the interference of the circulating currents, allowing the critical velocity, synchronization, and non-reciprocal response to be tuned through the rotation-induced phase. The combination of cavity-coupled stochastic Gross–Pitaevskii simulations and Bogoliubov–de Gennes analysis provides a unified description of the collective modes, transport thresholds, and cavity signatures of these dynamics. We have also demonstrated that the essential biased-transport physics can be realized using a purely Hermitian tilted optical-lattice potential. This provides an experimentally accessible alternative in which a controlled spatial energy bias produces source–sink population transfer and direction-dependent Josephson transport without requiring an explicitly complex potential. In the future, it would be interesting to explore Josephson dynamics in the vicinity of the exceptional point, where the imaginary eigenmodes begin to grow. Exploring changes in the current-phase relationships, current-voltage curves, and the Shapiro steps near the exceptional point would be interesting.  Another important aspect of our work is the connection between the underlying condensate dynamics and their nondestructive measurement. We show that the relevant Josephson and $\mathcal{PT}$-induced modes can be resolved through the power spectrum of the cavity output field, establishing cavity optomechanics as a highly sensitive probe of non-Hermitian matter-wave dynamics.

% Our results establish atomtronic SQUIDs as a versatile platform for exploring the interplay of non-Hermiticity, quantum interference, and driven Josephson dynamics. Several directions follow naturally from the present work. Experimentally, the predicted directional critical velocities, cavity spectral signatures, and asymmetric Shapiro steps could be investigated using dynamically controlled optical barriers and cavity-based homodyne detection. It would also be interesting to explore the behavior in the vicinity of the exceptional point, where the enhanced sensitivity and strong mode hybridization associated with non-Hermitian degeneracies may provide additional control of the Josephson response. The combination of barrier rotation and periodic driving offers a further route toward studying synchronization, geometric-phase effects, and more complex non-reciprocal transport. 

Beyond the mean-field description, incorporating quantum fluctuations and correlations could reveal the role of many-body effects in the stability and detection of the $\mathcal{PT}$ collective mode. Finally, extending the tilted-lattice realization to time-dependent and spatially programmable potentials may enable the design of reconfigurable atomtronic elements with controllable diode efficiency, frequency conversion, and synchronization. These developments could open new possibilities for using quantum fluids as platforms for non-reciprocal and non-Hermitian quantum technologies.

\section{Acknowledgments}
 M.B. thanks the Air Force Office of Scientific Research (FA9550-23-1-0259) for support. R.K. was supported by JSPS KAKENHI
(Grant No. 25K07190).  We also gratefully acknowledge our supercomputing facility Param-Ishan (IITG), where all the simulation runs were performed.

% \bibliography{citation.bib} 

\appendix

\counterwithin{figure}{section}

\section{Collective excitation spectrum}
\label{sec:BdG}
The dynamical behavior discussed in the Sec.~\ref{sec:I-V} can be understood in terms of the collective excitation spectrum of the condensate. To identify the relevant modes and determine the stability boundary of the $\mathcal{PT}$-symmetric SQUID, we perform a Bogoliubov-de Gennes (BdG) analysis about the stationary condensate state. We consider a small perturbation of the form
\begin{equation}
\Psi(\phi,t)=e^{-i\mu t/\hbar}
\left[
\psi_0(\phi)
+u_j(\phi)e^{-i\omega_j t}
+v_j^*(\phi)e^{i\omega_j^*t}
\right],
\label{eq:bdg_ansatz}
\end{equation}
where $\psi_0(\phi)$ corresponds to the ground state with final barrier configuration. Substituting Eq.~\eqref{eq:bdg_ansatz} into the time-dependent Gross-Pitaevskii equation (Eq.~\ref{Eq:bec}, with out the cavity coupling term with both the optical barriers and the complex $\mathcal{PT}$-symmetric potential) and retaining terms linear in the fluctuations yields the Bogoliubov-de Gennes (BdG) equations:
\begin{align}
\mathcal{L}u_j(\phi)
+g\psi_0^2(\phi)v_j(\phi)
&=\hbar\omega_j u_j(\phi),\\
-g\psi_0^{*2}(x)u_j(\phi)
-\mathcal{L}^*v_j(\phi)
&=\hbar\omega_j v_j(\phi),
\end{align}
where
\begin{equation}
\mathcal{L}=
-\frac{\hbar^2}{2mR_0^2}\frac{\partial^2}{\partial \phi^2}
+V(\phi)-\mu
+2g|\psi_0(\phi)|^2.
\end{equation}

The resulting eigenvalues are, in general, complex and may be written as 
%%%%%%%%%%%%%%%%%%%%%%%%%%%%%%%%%%%%%%%%%%%%%
\begin{equation}
\omega_j=\omega_j^{(R)}+i\omega_j^{(I)},
\end{equation}
where $\omega_j^{(R)}$ determines the oscillation frequency of the corresponding collective mode, while $\omega_j^{(I)}$ characterizes its growth or decay. In the dynamically stable regime, the relevant eigenvalues are predominantly real, corresponding to bounded collective oscillations. The emergence of a finite imaginary component signals the loss of dynamical stability associated with the non-Hermitian $\mathcal{PT}$-symmetric potential.

%-------------------------------------------------------
\begin{figure}[!htp]
\begin{center}
\includegraphics[width= 1.0\linewidth]{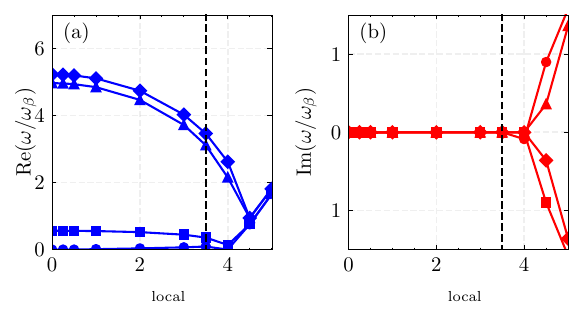}

\end{center}
\vspace{-12pt}
\caption{Bogoliubov-de Gennes (BdG) excitation spectrum for $V_0=17.5,\mu_0$, $\sigma=0.7,\zeta$, and $f_{\mathrm{bar}}=0$. (a) Real and (b) imaginary parts of the BdG eigenvalues as functions of $\Gamma_{\mathrm{local}}$. The coalescence of the real eigenvalues and the subsequent emergence of nonzero imaginary components identify the $\mathcal{PT}$-symmetry-breaking exceptional point, indicated by the black dashed line.}
\label{fig:figA1}
\end{figure}
 %-------------------------------------------------------
 %-------------------------------------------------------
\begin{figure}[!htp]
\begin{center}
\includegraphics[width= 1.0\linewidth]{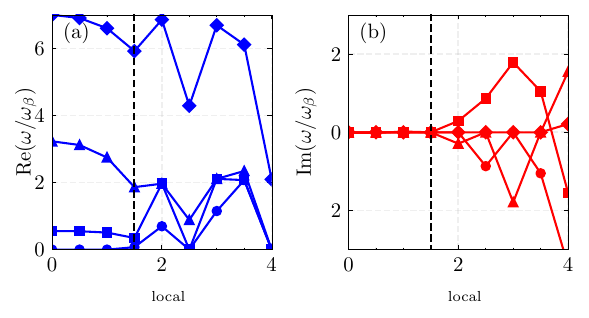} 
\end{center}
\vspace{-12pt}
\caption{BdG excitation spectrum for $f_{\mathrm{bar}}=0.9~\mathrm{Hz}$, $V_0=17.5,\mu_0$, and $\sigma=0.7,\zeta$. The ground state in this case is obtained at the final barrier position determined by $f_{\mathrm{bar}}=0.9~\mathrm{Hz}$ and $f_m = 0.0$ Hz. (a) Real and (b) imaginary parts of the BdG eigenvalues as functions of $\Gamma_{\mathrm{local}}$. The reduced exceptional-point threshold reflects the enhanced sensitivity to $\mathcal{PT}$-symmetry breaking induced by the increased asymmetry of the AQUID.}
\label{fig:figA2}
\end{figure}
 %-------------------------------------------------------
For $f_{\mathrm{bar}}=0$, the real and imaginary parts of the BdG eigenvalues are shown in Fig.\ref{fig:figA1}(a) and Fig.\ref{fig:figA1}(b), respectively. At weak non-Hermiticity, the collective modes possess real eigenfrequencies and can therefore be identified with well-defined oscillatory excitations of the condensate. As $\Gamma_{\mathrm{local}}$ is increased, the relevant real eigenmodes approach one another and their frequency separation decreases. At a critical value $\Gamma_{\mathrm{c}}$, the real eigenvalues coalesce. This coalescence defines the exceptional point of the $\mathcal{PT}$-symmetric system. Beyond this point, the eigenvalues cease to remain purely real and occur as complex-conjugate pairs, giving rise to nonzero imaginary components in Fig.~\ref{fig:figA1}(b). The appearance of these imaginary eigenvalues indicates the onset of exponential growth or decay of the corresponding collective excitations and hence the breakdown of the dynamically stable oscillatory regime.

The effect of the barrier-induced asymmetry on the excitation spectrum is further illustrated in Fig.~\ref{fig:figA2}, where the barrier configuration corresponds to $f_{\mathrm{bar}}=0.9~\mathrm{Hz}$. Compared with the symmetric configuration ($f_{\mathrm{bar}}=0$) in Fig.~\ref{fig:figA1}, the real eigenvalues approach coalescence at a lower value of $\Gamma_{\mathrm{local}}$, indicating a reduction in the exceptional-point threshold. Correspondingly, the onset of nonzero imaginary components occurs at a smaller gain--loss strength, as shown in Fig.~\ref{fig:figA2}(b). This shift demonstrates that the increased spatial asymmetry of the AQUID enhances the sensitivity of its collective excitation spectrum to $\mathcal{PT}$-symmetry breaking. Thus, even at the same gain--loss strength, the asymmetric barrier configuration can bring the system closer to the exceptional point and consequently modify its dynamical stability.

The BdG spectrum also clarifies the two characteristic frequency scales observed in the population-imbalance dynamics. The lowest collective mode corresponds predominantly to the Josephson plasma oscillation between the two barrier-defined wells. Its frequency is controlled by the effective tunneling coupling and is therefore sensitive to the optical-barrier height. The higher-frequency mode is associated with the density redistribution induced by the gain-loss structure of the $\mathcal{PT}$-symmetric potential. As discussed in Figs.~\ref{fig:fig1} and \ref{fig:fig2}, the latter mode is responsible for the additional oscillatory feature in the population imbalance and for the corresponding structure in the cavity-output spectrum. The BdG analysis thus provides a microscopic identification of the collective modes extracted from the dynamical simulations.

% \section{Detection of critical velocity}

%\clearpage
\bibliography{citation.bib} 

% \widetext

% \begin{center}
% \textbf{\large Supplementary material: Proposals for realizing a Josephson diode in Atomtronic circuits}
% \end{center}

\end{document}